\documentclass[manuscript]{aastex}
\def\lesssim{\mathrel{\hbox{\rlap{\hbox{\lower4pt\hbox{$\sim$}}}\hbox{$<$}}}}
\def\gtrsim{\mathrel{\hbox{\rlap{\hbox{\lower4pt\hbox{$\sim$}}}\hbox{$>$}}}}

\def\ltsima{$\;\buildrel < \over \sim \;$}
\def\simlt{\lower.5ex \hbox{\ltsima}}
\def\gtsima{$\;\buildrel > \over \sim \;$}
\def\simgt{\lower.5ex \hbox{\gtsima}}
\def\lesssim{\mathrel{\hbox{\rlap{\hbox{\lower4pt\hbox{$\sim$}}}\hbox{$<$}}}}
\def\gtrsim{\mathrel{\hbox{\rlap{\hbox{\lower4pt\hbox{$\sim$}}}\hbox{$>$}}}}
\def\gtrless{\mathrel{\hbox{\rlap{\hbox{\lower4pt\hbox{$<$}}}\hbox{$>$}}}}
\def\rightleftharpoons{\mathrel{\hbox{\rlap{\hbox{\raise2pt\hbox{$\rightharpoonup$}}}\hbox{$\leftharpoondown$}}}}
\def\notrightleftharpoons{\mathrel{\hbox{\rlap{\hbox{\raise1.5pt\hbox{$\;\mid$}}}\hbox{$\rightleftharpoons$}}}}
\def\dbar{\mathrel{\hbox{\rlap{\hbox{\raise3pt\hbox{$-$}}}\hbox{$d$}}}}
\def\hbar{\mathrel{\hbox{\rlap{\hbox{\raise3pt\hbox{$-$}}}\hbox{$h$}}}}
\def\nubar{\mathrel{\hbox{\rlap{\hbox{\raise2pt\hbox{$-$}}}\hbox{$\nu$}}}}
\def\lambdabar{\mathrel{\hbox{\rlap{\hbox{\raise2pt\hbox{$-$}}}\hbox{$\lambda$}}}}
\def\BbbV{\mathrel{\hbox{\rlap{\hbox{\raise2.5pt\hbox{${\rm v}$}}}\hbox{${\rm V}$}}}}
\def\BbbT{\mathrel{\hbox{\rlap{\hbox{\raise2pt\hbox{${\rm T}$}}}\hbox{${\rm T}$}}}}

\def\dddot{\hbox{\rlap{\hbox{\raise 8pt\hbox{${\bf ...}$}}}\hbox{$$}}}

\def\ltsima{$\;\buildrel < \over \sim \;$}
\def\simlt{\lower.5ex \hbox{\ltsima}}
\def\gtsima{$\;\buildrel > \over \sim \;$}
\def\simgt{\lower.5ex \hbox{\gtsima}}

\shorttitle{Galaxy Mass, Metallicity, Radius and SFRs}
\shortauthors{Brisbin and Harwit}

\begin{document}

\title{Galaxy Mass, Metallicity, Radius and Star Formation Rates}
\author{Drew Brisbin\altaffilmark{1} and Martin Harwit\altaffilmark{1,2}}

\email{brisbind@astro.cornell.edu}

\altaffiltext{1}{Center for Radiophysics and Space Research, Cornell University,Ithaca, NY, 14853}
\altaffiltext{2}{511 H street, SW, Washington, DC 20024-2725}

\begin{abstract}
Working with 108,786 Sloan Digital Sky Survey low redshift galaxies we have examined the relation between galaxy mass, metallicity, radius, and star formation rates primarily in the central portions of galaxies.  We subdivided the redshift range covered in our sample, $0.07\leq z\leq0.3$, into three narrower redshift bins, and three sets of radial size.  We show that for 72\% of the galaxies the observed gas metallicities, $Z_x$, are consistent with (i) a quantitative physical relation for star formation through episodic infall of gas of metallicity $Z_i = 0.125\times 10^{-3}\pm 1.25\times 10^{-3}$; (ii) thorough mixing of infalling and native gas before onset of star formation; (iii) a star formation rate (SFR) proportional to the 3/2 power of the infalling mass rate, $\dot M_i$; and (iv) intermittent quiescent phases devoid of star formation during which the native gas in a galaxy exhibits a characteristic elevated gas metallicity, $Z_0$, dependent on galaxy mass, $M_*$, and a characteristic ratio of stellar mass to native mass of gas, $M_g$.   Most if not all our star-forming galaxies with  $M_* \leq 2.0\times 10^{10}M_{\odot}$, and many with $M_* \geq 2.0\times 10^{10}M_{\odot}$ and large radii appear fed by infall.  Smaller massive galaxies with high $Z_x$ and high star formation rates show more complex behavior.  A mean-field-theory toy model for the physics of infall  accounts for the $(SFR)\propto \dot M_i^{3/2}$ relation and permits us to estimate the mean densities and velocities of clumps of baryonic matter traversing the dark matter halos in which the SDSS galaxies may be embedded. 
\end{abstract}


\section{Introduction}

The causes and conditions surrounding star formation appear to be highly complex. Despite many attempts to fully understand the different processes that may be involved, a globally applicable theory still appears lacking.  The past few years, have witnessed many attempts to account for star formation laws, such as the Schmidt-Kennicutt law, or extensions of them \citep{mad2010,shi2011}.  Until recently, and even now, star formation rates have often been treated largely in terms of observable or inferred physical conditions within well-isolated galaxies.\citep{cal2009,kru2012}  Within the past year or two, however, chemical considerations have begun to invoke infall and/or outflow of matter, some of which may have observable consequences on star formation only after considerable delay. \citep{kob2007,bro2007,dav2011} 

Recently published observational evidence on correlations between star formation rates (SFRs) and gas metallicity as a function of galaxy stellar mass is certainly striking \citep{man2010}.  It suggests that star formation may need to be considered in entirely new ways that emphasize infall of matter from enveloping halos.

The Sloan Digital Sky Survey, SDSS, has made available star formation rates and gas metallicities for more than a hundred thousand galaxies at red shifts $0.07\lesssim z\lesssim 0.3$.  \citep {tre2004,gal2005,ell2008,kew2008,man2010,pen2010,pen2011,aih2011,yat2011}.  Although these data are largely statistical and may lack the more detailed correlations that observations on star formation in nearby galaxies make possible, the advantage of working with SDSS data is the ability to establish commonalities in the properties of tens of thousands of galaxies selected from a homogeneously acquired set.  In particular, \citet{ell2008} noted a number of striking correlations between galaxy mass, radius, and metallicity, as well as between mass, star formation rate (SFR) or specific star formation rate (SSFR), and metallicity.  This was complemented by a finding of \citet{man2010} who produced a  three-dimensional Cartesian plot having galaxy mass, star formation rate, and metallicity as its orthogonal coordinates.  This clearly shows the population of star-forming galaxies out to redshifts at  least as high as $z \sim 0.3$ lying on a smoothly curved surface. The range of metallicities at each point on this surface is narrow, revealing the surface to be a thin curved sheet.  Recent work by \citet{yat2011} analyzes the findings of \citet{man2010} in greater detail.

The persuasive correlations these studies have revealed between high star formation rates and low metallicity, have made a compelling case for considering infall of pristine, low- or zero-metallicity gas as a primary factor governing massive star formation.  

That infall of high-velocity clouds is prevalent at least in the Galaxy has recently been convincingly documented by  \citet{leh2011} through observations with the Hubble Space telescope. They identified the infalling gases through their ionized oxygen, silicon and carbon absorption lines seen against stars at high Galactic latitudes, and estimated the rate of infalling mass to possibly be of the order of 0.5 $M_{\odot}$ yr$^{-1}$. This is potentially sufficient to maintain star formation at an expected rate for the Galaxy.   Two questions, however, remain unanswered. The observed infall velocities are of order 100 km s$^{-1}$, which may be sufficiently low for the material to simply be gas that is falling back into the galaxy, having been previously ejected to some appreciable height.  Further, the assumed metallicity of the infalling gas, estimated to be $Z\sim 0.2 Z_{\odot}$, leaves unanswered the question of whether this could be the metallicity of infalling gas currently traversing the local halo, or again just reflect the metallicity of previously ejected material now falling back into the Galaxy.

These questions have been given impetus by the work of \citet{dav2011} and \citet{dav2011b}, who have emphasized the importance of supernova ejection of high-metallicity material, which then may fall back into a galaxy, depending on whether or not the ejection powers escape velocities. 

Evidence for resolving these questions may soon accumulate through observations of outflows from star-bursting galaxies, as shown by recent observations on PG1206+459 by \citet{tri2011}. While uncertainties about the metallicity of the outflowing gas and the extent to which some of this gas may fall back into the observed galaxy still prevail, observations of the attempted kind may soon provide informative results.  

Our motivation in pursuing the analyses provided in the present paper has been to unambiguously quantify the relationship between infall of gas and star formation rates.  The data of \citet{man2010} strongly suggest that the correlation between SFR and low metallicity requires the infall of low-metallicity or pristine extragalactic gas.  But other explanations for this correlation, for example those of \citet{cow2011}, have not be ruled out.  In these circumstances, it appeared to us that the sheer wealth of data provided by the SDSS might serve as a guide, which would permit a more compelling analysis.  

Obtaining an empirical relationship of the type we exhibit in the following sections, however, is not enough.  Ideally one would like to have a physical model that accounts for an observed relation, is consistent with other observational data and, where possible with extant theory.  For this purpose, we have developed a mean-field-theory toy model that indicates how the metallicity and velocity of infalling gas may be related to SFR and to the native gas content of a galaxy as discussed, for example, by \citet{tre2004}; \citet{dal2007}; or \citet{pee2011}.  Such a model should also take into account recent calculations on infall of extragalactic gas into galaxy halos, to the extent that pristine or low-metallicity gases are most likely to have accumulated in a halo.  Moreover, the velocity at which this matter traverses the halo and impacts on a galaxy should still reflect that history.

Although a toy model cannot be as comprehensive as dedicated computer simulations developed, e.g., by \citet{bro2007}; \citet{kob2007}; or \citet{kly2011}, it provides ready insight on the physical relations between different parameters, and allows observers to check for potential confirmation or refutation of hypotheses on the basis of readily apparent scaling relations.  Within these restrictions, the toy model we develop in the present paper appears at least in rough agreement with available data, and makes specific assertions  that may be negated or verified by future observations.

A key feature of the present paper is a quantitative demonstration that most, if not all star-forming galaxies with  $M_* < 2.0\times 10^{10}M_{\odot}$, and many with $M_* \geq 2.0\times 10^{10}M_{\odot}$ and large radii appear to be fed by infall of pristine or low-metallicity gas.  More massive galaxies, however, appear to form stars  also by other means, possibly through consumption of gas injected into interstellar space by outflow from low-mass evolved stars, as suggested by \citet{lei2011}.  The evidence we provide is  chemical and based on a novel quantitative relationship we establish between SFR, metallicity and a galaxy's stellar mass.

The wealth of galaxies observed in the SDSS permits our toy model to examine not only the correlations between such properties as galaxy mass, radius, SFR, and metallicity, but, to a limited extent, also the evolution of these quantities with redshift.  We restrict ourselves to galaxies observed in the redshift range $z = 0.07$ to 0.3, corresponding to an interval of about 2.5 Gyr spanning an epoch that began 3.4 and ended 0.9 Gyr ago. We assume standard flat $\Lambda$CDM cosmology, $\Omega_{CM} \sim 0.27,\ \Omega_{DE}\sim 0.73$, and a Hubble constant, $H_0 = 71$ km s$^{_1}$ Mpc$^{-1}$ \citep{wri2006}.

We structure our paper as follows: Section 2 details our adopted data processing methods.  Section 3 characterizes observational results.  Section 4 introduces our main finding, namely that the relation between metallicity and SFRs in galaxies of fixed mass and radius is quantitatively consistent with dilution of native gas in these galaxies by infalling matter.  Section 5 discusses our findings in terms of a mean-field-theory toy model that clarifies and quantitatively accounts for the observations. On the basis of the SDSS observations, section 6 derives mean values for the velocities $\langle V\rangle$ and densities $\langle \rho\rangle$ of infalling matter.  In galactic halo models, these appear to be associated with mean properties of clumps of gas traversing the halos. In Section 7, we briefly discuss the plausibly permitted range of differences between the mass infall rates that trigger star formation and the resulting star formation rates.
A final section 8 discusses our findings in terms of other recently published studies. 

\section{Methods}

Our selection of galaxies comes from the MPA-JHU catalog compiled from SDSS DR7 available online (http://www.mpa-garching.mpg.de/SDSS/DR7/).  

To permit closer comparison of our data and those of \citet{man2010}, we have used identical selections in redshift, (between z=0.07 and 0.3), H$\alpha$ signal to noise ratio ($\geq$25), and we similarly excluded active galactic nuclei, AGNs, using the criteria put forward by \citet{kau2003}.  We have also followed identical procedures to find stellar mass, SFRs, and metallicities.  Our stellar masses are taken directly from the MPA-JHU catalog as inspired by \citet{kau2003_b}, with a correction to convert from masses based on a Kroupa IMF \citep{kro2001} to a Chabrier IMF \citep{cha2003}.  Star formation rates are determined using H$\alpha$ and are based on the work of \citet{ken1998} with a correction for a Chabrier IMF.  

While H$\alpha$ is subject to foreground dust extinction, we correct for this using the Balmer decrement, along the same lines as \citet{car1989}.  We excluded any sources with large extinction corrections corresponding to A$_V > 2.5$.  Metallicities are determined from the strong line diagnostics $R_{23}=([OII]3727+[OIII]4958+[OIII]5007)/H\beta$ and $[NII]6584/H\alpha$ as presented in \citet{nag2006} and \citet{mai2008}.  We note that other alternatives for obtaining metallicities have recently been proposed by \citet{lar2010}; but for ease of comparison we have opted to follow procedures adopted by \citet{man2010} and \citet{ell2008}.

The bulk of our selected sources cover a galactic stellar mass range from $\sim 10^9$ to $2\times 10^{11} M_{\odot}$, with star-formation rates ranging roughly from $0.035$ to $6.4 M_{\odot}$ yr$^{-1}$, and oxygen abundances from $\sim4\times 10^{-4}$ to $1.4\times 10^{-3}$, straddling a Solar System abundance of $\sim 8.5 \times 10^{-4}$.  Because we concentrate on the additive properties of metallicity, rather than metallicity ratios, we express metallicities throughout in terms of actual oxygen abundances, as opposed to their logarithmic values more conventionally adopted. 

The redshift range covered both by \citet{man2010} and us is set at a minimum value of $z = 0.07$ to ensure that the [OII]$\lambda3727$ emission line is well within the useful spectral range and that the $3"$ aperture of the SDSS spectroscopic fiber will sample a significant fraction of a galaxy's surface area.  At $z = 0.07$, an aperture of $3"$ corresponds to an angular diameter of 4 kpc implying that we probe only the central 2 kpc regions of a galaxy.  At $z = 0.3$, our observations probe galaxies out to radial distances  $\sim6.6$ kpc, and thus sample the galaxies more fully \citep{wri2006}.

In addition to these selection criteria imposed by \citet{man2010} and ourselves, we required our sample to have an observed Petrosian half-light radius in the {{$r$-color}} band, r$_{50}$.  This is required, in the spirit of \citet{ell2008}, to investigate the role galactic radius plays in star formation and metallicity.  We have also taken pains to eliminate duplicate observations from our SDSS sample.  Where an object was observed multiple times, we averaged its properties across the multiple entries.

Both \citet{man2010} and \citet{ell2008} have treated galaxies in the redshift range we cover as though they were coeval.  To examine whether or not further insight could be gained through study of the provenance of their galaxies, we have divided the SDSS galaxies into individual tables corresponding to three separate redshift ranges, $0.07\leq z \leq 0.10,\ 0.10\leq z\leq 0.15,\ {\rm and}\  0.15\leq z\leq 0.30$; and three Petrosian half-light radii r$_{50}$ ranges, small r$_{50} < 3.74$ kpc, medium 3.74$\leq$r$_{50} < 5.01$, or large r$_{50} \geq $5.01. Each table divides galaxies according to logarithmic mass and logarithmic SFR. These tables all are available as part of our paper online.

  Although many of the entries in our tables show galaxy populations that may be quite sparse in certain galaxy mass/SFR bins, we considered our findings significant only if based solely on  bins that have $\geq 50$ galaxies per bin.

We are aware that \citet{cha2005} has updated his initial mass function and that it is in better agreement with current data as indicated by \citet{par2011}.  However, in order to compare our observational data and analytic model with those of \citet{man2010} and \citet{ell2008}, in section 6 below, we have chosen to adopt an initial mass function identical to theirs in order to maintain our respective data sets as homogeneous as possible, so that a more direct comparison of our findings and theirs may emerge.  This procedure admittedly maintains inherent biases that all three data sets may propagate, but enables a more meaningful comparison of results.

While the SDSS provides a way of studying star formation statistically in a sample exceeding a hundred thousand galaxies, both our selection and those of  \citet{man2010} and \citet{ell2008} restrict themselves to observations of the central portions of galaxies, in many of which spiral arms no longer are well defined.  Our sample thus is not adequate to addressing questions of spiral structure or its effects on star formation,  and generally obstructs direct comparison to work studying star formation over larger portions of galaxies' disks.  Nevertheless, given the importance of processes taking place in the nuclear portions of galaxies, our choice of regions has a compelling interest of its own.

\section{The Observed Galaxy Sample}

\begin{figure}[h!]
	\centering
		\includegraphics[width=1\textwidth, trim= 7inches 1.5inches 7inches .7inches]{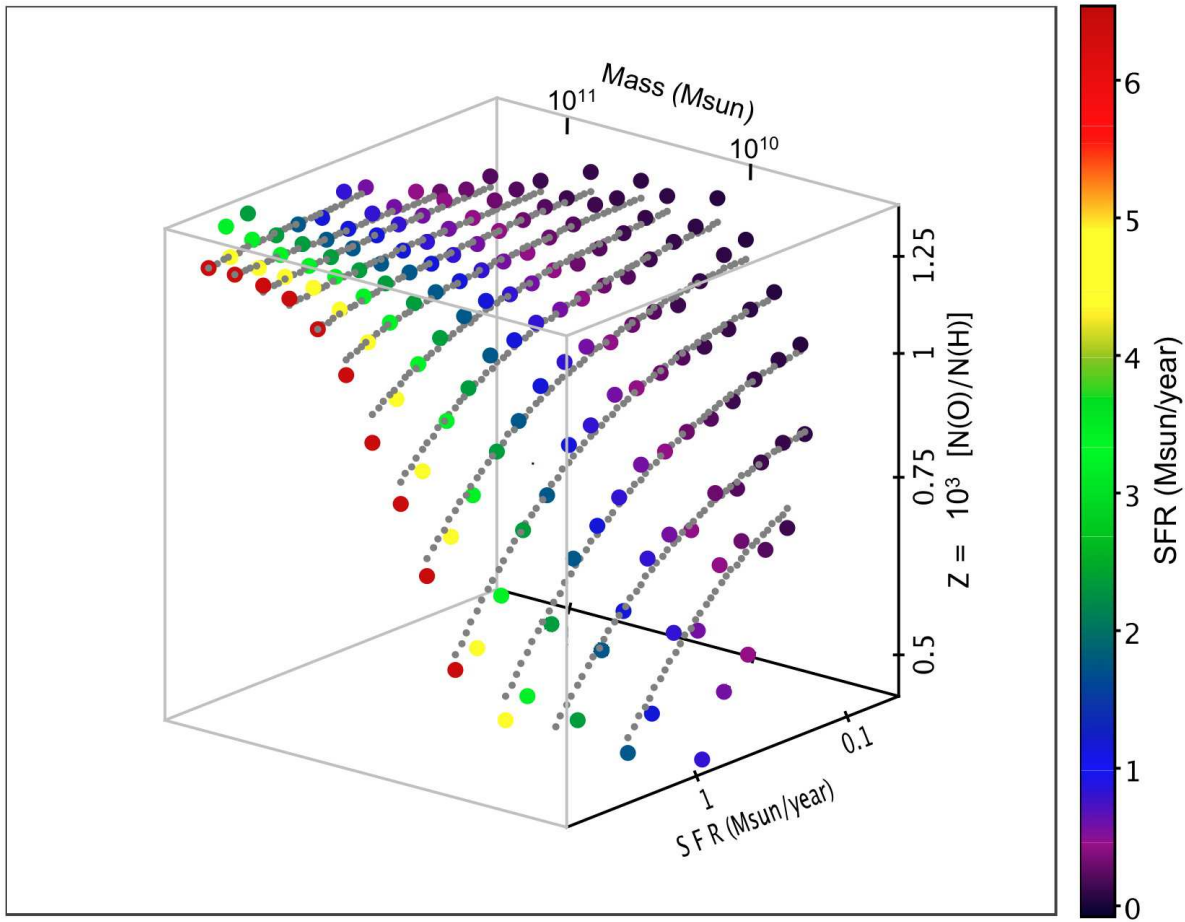}
	\caption{Median metallicity for samples across our range of mass and SFR.  This includes samples across all redshift cuts and all radius cuts. Here we include only samples with at least 50 sources at a given mass and SFR.  The small grey points indicate toy model fits to our data, as detailed in section 5.  Complete numerical tables of these data as well as subsets in redshift and galactic radius are available online.}
	\label{fig:table1}
\end{figure}

In Figure \ref{fig:table1} we present our galaxy sample for the entire redshift and radial size range divided into logarithmic SFR and stellar mass bins.  In each bin where $>$50 sources exist the metallicity is plotted on the vertical axis.

A full complement of nine sets of tabulations showing galaxy stellar mass, SFR, gas metallicity and metallicity scatter, separately for each of the three redshift intervals and each of the three ranges of Petrosian radii, is provided online. The tables provide much of the statistical information of interest to our paper derivable from the SDSS.

To detect whether evolutionary effects might have had an impact on previously published findings we divided the available data into the three redshift ranges defined above.   In mass/SFR bins where there were sufficient data in all three redshift epochs our tables provide data on the evolution of metallicity with time.  There are $\sim 40$ overlapping mass/(SFR) bins in the three  red-shift tables, in which each bin is populated by at least 50 galaxies. In all, this comprises several thousand galaxies in each redshift range.  Strikingly, galaxies with identical mass and SFR bins have median metallicities that vary across redshift by at most a few percent, somewhat randomly and well within standard deviations that generally are of order 15\%. This is in satisfactory accord with the findings of \citet{sav2005}, who observed galaxies in the somewhat higher redshift range, $0.4\lesssim z\lesssim 1.0$ and found no significant evolutionary trends in the galaxy mass-metallicity relation.

Evidently, galaxies with identical masses and star formation rates, in the mass range from  $\sim 7\times 10^9$ to $\sim 10^{11} M_{\odot}$,  appear to also share other significant properties despite the $\sim 2.5$ Gyr epoch spanning the redshift range $0.07\leq z \leq 0.30$.   In  this respect, the metallicities exhibited in the single table that \citet{man2010} provided appear to be robust even though they included data on galaxies over the entire $z = 0.07\ -\ 0.30$ redshift range and thus galaxies observed at epochs that may differ by as much as 2.5 Gyr within individual bins. 

\begin{table}[t]
\caption{Main Sequence Galaxies:  Peak Populations in the Mass/SFR Tables}
\vskip0pt\vskip-6pt\vskip0pt
{\vskip-6pt}
\begin{center}
\begin{tabular}{lccc} \hline\noalign{\vskip3pt}
Redshift Coverage of Table&$0.3\geq z\geq 0.15$&$0.15\geq z \geq 0.10$&$0.10 \geq z \geq 0.07$\\
\noalign{\vskip3pt}\hline\noalign{\vskip3pt}
Metallicity Range Spanning all Masses&&&\\
(N(O)/N(H) $ \times 10^{3}$)& 0.61 - 1.37& 0.51 - 1.39&0.473 - 1.34 \\
Galaxy Mass, $M_*$, in Peak Population Bin&&\\
in units of $10^{10} M_{\odot}$&5.62 & 2.82 & 1.41\\
Metallicity in Peak Population Bin&&&\\
(N(O)/N(H) $ \times 10^{3}$)&1.26&1.23&1.17\\
SFR in Peak Population Bin&&&\\
(in units of $M_{\odot}$ yr$^{-1}$) &2.239&1.122 &0.398\\
$\tau\equiv M_*$/SFR at Peak Population&&&\\
(in units of $10^{10}$ yr)&2.5& 2.5 & 3.5\\
\\\noalign{\vskip3pt}\hline
\end{tabular}
\end{center}
\end{table} 
 
We find that even at the highest SFRs, the galaxies in our sample could not have gained sufficient mass during the $\sim 2.5 $ Gyr between epochs $z = 0.3$ and 0.07 to shift significant numbers of galaxies from one mass column into a neighboring column.  As table 1 indicates, SFRs diminish toward lower galaxy mass but in such a way as to apparently enable all of the galaxies to accumulate mass at a rate proportional to their own masses. In effect, the specific star formation rate (SSFR), defined as the star formation rate per galaxy mass, remains constant for this peak population component.  \citet{pen2010} have also noted that the deviation from constancy of the SSFR is quite small across all masses in the SDSS.  Here we are finding a related result, namely that their relation holds even though a significant number of higher-mass galaxies are observed at an epoch 2.5 Gyr earlier than virtually all low-mass galaxies.  The time $\tau_d$ galaxies at all masses would require to double their mass at their observed SFRs is practically constant.   $\tau_d$ is the reciprocal of SSFR, and assumes values of  $\tau_d \sim 2.5 - 3.5\times 10^{10}$ yr, roughly 2 times the age of the Universe at each of the observed epochs \citep{wri2006}.

These data should, however, not be interpreted as indicating low, steady infall rates that maintain a galaxy in a relatively stable state of mass accretion.  We envisage an individual galaxy of given mass periodically moving down or up within its mass column in one of our tables, as infall from its surrounding halo increases or decreases both the infall and the star formation rate within the galaxy.  As we  begin to show, starting with the next section, mass infall rates into galaxies vary across a range spanning nearly two orders of magnitude, need not be constant in time, and appear to be consistent with episodic star formation in the SDSS sample we have studied.  

\section{Infall and Metallicity Dilution of Native Gas} 

We now examine the extent to which the amount of mass $\dot M_i\tau_i$, which has fallen into  a galaxy before onset of star formation, dilutes the initial metallicity $Z_0$ of the mass of native gas,  $M_g$, present before the onset of star formation.

If we assume that the infalling matter has metallicity $Z_i$, and mixes thoroughly with the galaxy's native gas, we expect that the observed metallicity $Z_x$ should be given by the ratio of the aggregated mass in metals after infall, divided by the total aggregated gas.
\begin{equation}
Z_x = \frac{(\dot M_i\tau Z_i + M_gZ_0)}{(\dot M_i\tau + M_g)} 
\label{eq:zxmass1}
\end{equation}
From equation (\ref{eq:zxmass1}) we can see that in galaxies which have an amount of infalling gas equal to the native gas mass, $\dot M_i\tau=M_g$, the  observed metallicity will be half the native metallicity, $Z_x = \frac{1}{2}Z_0$,  if the infalling gas has zero metallicity, $Z_i = 0$.  

To test equation (\ref{eq:zxmass1}) broadly, we adopt a trial function expressing mass infall rates, $\dot M_i$, for which we lack observations, in terms of SFRs for which we do have observations,
\begin{equation}
\dot M_i = \alpha (SFR)^{2/3}, \ {\rm equivalent\ to}\  (SFR) = \epsilon \dot M_i^{3/2}\ .
\label{eq:crux}
\end{equation}

Here $\alpha$ and $\epsilon \equiv \alpha^{-3/2}$ are constants of proportionality having some as-yet-unknown value but units, respectively of $M_{\odot}^{1/3}$ yr$^{-1/3}$ and $M_{\odot}^{-1/2}$ yr$^{1/2}$. 

We will justify the $\dot M_i$ - SFR relations of equation ({\ref{eq:crux}) in section 5. The second of these relationships is the more physical.  The factor $\epsilon$ is a measure of star formation efficiency.  But, for now, we will make use of the relationship based on $\alpha$, because 
(SFR) is an observationally determined quantity, whereas $\dot M_i$ is derived.
Using this relation the observed gas metallicity takes the form 
\begin{equation}
Z_x = \frac {Z_0 \left[1+\frac{Z_i}{Z_0}\left(\frac{\alpha(SFR)^{2/3}\tau_i}{M_g}\right)\right]}{\left( 1 + \frac{\alpha(SFR)^{2/3}\tau_i}{M_g}\right)}\ . 
\label{eq:zxsfr}
\end{equation}
 
\begin{table}[t]
\caption{Star Formation Rates and Observed and Computed Metallicities for low redshift, small half light radius galaxies with M$_*=$0.5$\times$10$^{10}$M$_\odot$ in Figure \ref{fig:sampledilution}.}
\vskip0pt\vskip-6pt\vskip0pt
{\vskip-6pt}
\begin{center}
\begin{tabular}{lccc} \hline\noalign{\vskip3pt}
(SFR)	&	$(SFR)^{2/3}$	&	10$^3\times Z_x$ Observed	&	10$^3\times Z_x$ Modeled	\\
M$_\odot$/yr& 				&N(O)/N(H)						& 	(Eq \ref{eq:zxsfr})	\\
\noalign{\vskip3pt}\hline\noalign{\vskip3pt}
0.10  	&	0.22  	&	1.00  	&	0.99  	\\
0.14  	&	0.27  	&	0.97  	&	0.97  	\\
0.20  	&	0.34  	&	0.94  	&	0.95  	\\
0.29  	&	0.43  	&	0.90  	&	0.92  	\\
0.40  	&	0.55  	&	0.88  	&	0.89  	\\
0.57  	&	0.69  	&	0.86  	&	0.86  	\\
0.81  	&	0.87  	&	0.82  	&	0.82  	\\
1.14  	&	1.09  	&	0.79  	&	0.77  	\\
1.61  	&	1.37  	&	0.72  	&	0.72  	
\\\noalign{\vskip3pt}\hline
\end{tabular}

\end{center}
\label{tab:sampledilution}
\end{table} 
\newpage
\begin{figure}[h!]
	\centering
		\includegraphics[height=1\textwidth,angle=90]{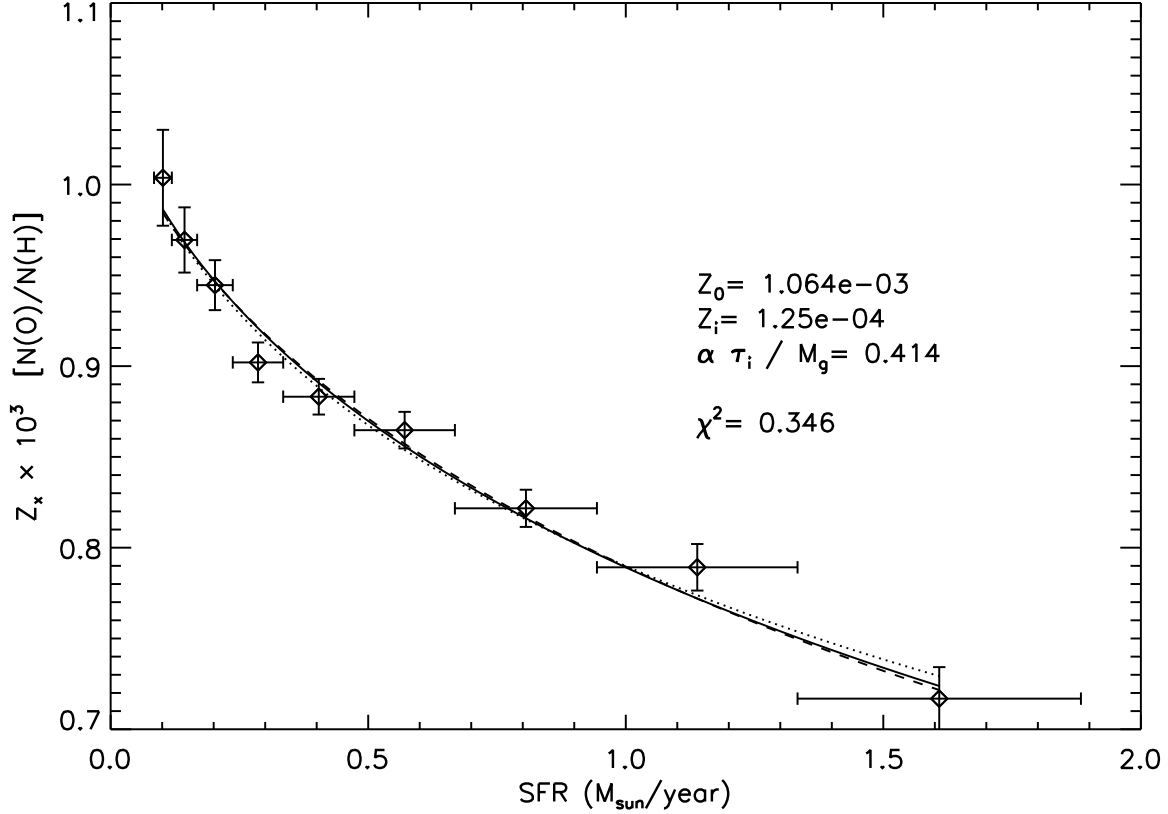}
	\caption{Observed gas metallicities at various SFRs for $M_*=0.5\times10^{10}M_\odot$ galaxies with low redshifts and small half light radii.  The fitted equation (\ref{eq:zxsfr}) with $Z_0=1.064\times10^{-3} N(O)/N(H)$, $(\tau_i \alpha)/M_g=0.414$ (M$_\odot/$year)$^{-2/3}$, and $Z_i=0.125\times10^{-3}$ is plotted as a solid line.  The fit is acceptable with a $\chi^2$ value of 0.35.  Dashed and dotted lines indicate fits based on our lower and rough upper limits on $Z_i$, 0 and $0.25\times10^{-3}$ respectively. Similar plots for galaxies with different stellar masses, radii and redshifts are available online.}
	\label{fig:sampledilution}
\end{figure}

As an example, we may use the entries for low redshift ($0.07\leq z \leq0.10$,) medium radius ($3.74 \leq r_{50} \leq 5.01$ kpc,) galaxies of mass $0.5\times10^{10}M_{\odot}$  from our tables.  Adopting the observed metallicities in units of oxygen abundances (10$^3\times$N(O)/N(H)), at the available SFRs $(M_\odot$ yr$^{-1}$), we can calculate best fit values of $Z_0$ and $(\tau_i \alpha)/M_g$. For these data we find a good fit with Z$_0$=1.06$\times$10$^{-3}$ N(O)/N(H) and $(\tau_i \alpha)/M_g=0.414$ (M$_\odot/$year)$^{-2/3}$. Table \ref{tab:sampledilution} shows the results of this sample calculation and Figure \ref{fig:sampledilution} shows equation (\ref{eq:zxsfr}) fitted to the sample data.  Regrettably, we cannot isolate $M_g$ from $\alpha \tau_i$ with the data given, and can derive solely their ratio $(\alpha \tau_i)/M_g$.

Online we provide our metallicity fits like those of Figure \ref{fig:sampledilution} for all stellar masses, radii, and redshifts for which we have adequate statistics --- by which we mean $>$ 50 galaxies per mass/SFR bin and 5 of such bins per galaxy mass column.   For each sample of galaxies we performed a $\chi^2$ analysis, testing equation (\ref{eq:crux}) with a range of $0.1\leq Z_0\times10^3\leq1.5$ N(O)/N(H) and $0\leq\frac{\alpha \tau_i}{M_g}\leq5$ (M$_\odot/$year)$^{-2/3}$ and selecting the lowest reduced $\chi^2$ fit.  Observational uncertainties in the emission lines used in our strong line diagnostics are propagated through to find the uncertainty in metallicity for each mass bin.  Horizontal error bars represent the uncertainty in SFR due to the SFR bin width.  To make use of the SFR uncertainties in our $\chi^2$ analysis we found how much the model metallicity varies over the range of each SFR bin and took half this variance to add in quadrature with the metallicity uncertainty.  This conversion of uncertainty in SFR to uncertainty in metallicity is performed for each SFR bin in each galaxy sample with every trial of model parameters.

While we expect $Z_0$ and $\frac{\alpha \tau_i}{M_g}$ to vary between galaxies of different mass, in our naive picture the infalling gas should not anticipate what kind of galaxy it is falling into and thus should not vary systematically between mass bins. As \citet{dav2011} and \citet{dav2011b} have pointed out, however, feedback in the form of supernova ejection may pollute extragalactic gas, raising the possibility that infalling gas might already have significant metallicity. For that reason, in our $\chi^2$ analysis we allow the infalling gas metallicity, $Z_i$, to be nonzero but fix it at the same value for each sample. Repeating the whole fitting procedure many times with different fixed values of $Z_i$ we find that a value of $Z_i=0.125\times10^{-3}$ minimizes the total $\chi^2$ of our combined fits.  A range of low metallicity solutions, spanning $0\leq Z_i\leq0.20\times10^{-3}$ 
provide nearly equally good fits, while for $Z_i\geq0.30\times10^{-3}$ the fits become clearly worse and the total $\chi^2$ value sharply increases.  
We therefore take $0.125\times10^{-3}$ 
as our infalling gas metallicity with the caveat that our results are consistent with any very low metallicity infalling gas  $Z_i\sim 0.125\times 10^{-3}\pm 0.125\times 10^{-3}$.

Our metallicity fits show that 72\% of the SDSS galaxy samples obey the metallicity / star-formation relations and are well fit by equation (\ref{eq:zxsfr}).  Given our small number of degrees of freedom (which ranges between 3 and 13) for each of our metal dilution samples, we used a reduced $\chi^2$ value of 2 for our cutoff.

For the curves that provide good fits, we averaged both the $Z_0$ and $M_g$ values obtained for each group of galaxies with a given stellar mass $M_*$.  For the mass sequence from $M_* = 0.36$ to $11.2 \times 10^{10}M_{\odot}$ for which we have sufficient data, we find a monotonic increase in metallicity going from $Z_0$ = 0.90 to $1.37\times 10^3$ N(O)/N(H).  We normalize the other fitting parameter by multiplying it by stellar mass, $\alpha \tau_i M_*/M_g$.  This parameter stays relatively constant over the entire galaxy mass range, hovering around an average value of (2.3 $\pm$ 0.7) $\times$10$^9$ (yr$^2$ M$_\odot$)$^{1/3}$.  On the assumption that $\alpha$ and $\tau_i$ are constants across our sample bins, this parameter is indicative of the stellar to gas mass ratio. If the gas fraction in a typical galaxy were $M_g/M_*\sim x$, the corresponding fitted parameter would be $\alpha \tau_i\sim 2.3\times 10^9x$ (yr$^2$ M$_\odot$)$^{1/3}$.

Sets of galaxies not obeying equations (\ref{eq:zxmass1}) and (\ref{eq:zxsfr}), invariably exhibit high metallicities at high SFRs and masses $M_*\geq 2.0\times 10^{10}M_{\odot}$, suggesting that for these galaxies star formation independent of pristine or low-metallicity mass infall plays a significant role.

The good quality of the fits we observe convince us that the dilution we have modeled actually takes place.  Particularly interesting then is that the anticipated dilution must take place before star formation is initiated, so that the observed HII regions, which enable us to determine SFRs and metallicity values, reflect the conditions that triggered the formation of the stars. 

Two points may still be noted.  \citet{tre2004}, \citet{dal2007}, and \citet{pee2011} have evaluated a number of trends of gas mass and metallicity in galaxies with different stellar masses.  However, as the detailed atlas of galaxies compiled by \citet{ler2008} makes clear, global ratios of gas-to-stellar mass vary enormously, depending on whether one is dealing with a narrowly defined central region of a galaxy, as defined by the SDSS spectroscopic aperture used throughout the present paper, or entire galaxies, as studied by \citet{ler2008}.  Extended galaxies generally show the ratio of gas to stellar mass in a galaxy's central regions to be of the order of a few percent, even when that ratio, averaged over the entire galaxy is one or two magnitudes higher. But even in these central regions, the ratio $M_g/M_*$ can vary greatly.

For this reason, legitimate questions may be raised about our assumption that some well-defined metallicity and fractional amount of native gas can be defined for galaxies in a given stellar mass range, and that both parameters may be expected to vary systematically and continuously over galaxies having different stellar masses. In particular, one may ask what physical basis might account for such a relationship.  

The response to these questions is that, in the present investigation, both the gas fractions and the metallicities involved are statistically determined. For each of the galaxy masses and SFRs considered we deal with groups no smaller than 50 galaxies. For galaxies in each stellar mass range, both the derived native gas metallicity and an assumed gas fraction derived from all the distinct SFR ranges are averaged over hundreds if not thousands of galaxies.  To the extent that the SDSS galaxies fall into different classes, governed by distinct physical conditions, there may be no single physical reason for assuming that such average values actually exist, other than that the occurrence of each class of galaxy in the general distribution of galaxies follows some well-defined evolutionary history, and that the physical parameters governing these galaxies' gaseous components fall into a relatively restricted range.

Table \ref{tab:Z0summary} provides additional insight offered by an analysis of galaxies of different radii.  In this table we show the fitted native gas metallicity values, Z$_0$, at all masses and radii where adequate statistics were available in a mass range spanning 0.25 - 4 $\times$10$^{10}$ M$_\odot$.  Note that satisfactory fits using equation (\ref{eq:zxsfr}) were found for all three radii, at low and medium redshifts at galaxy masses 0.7 $\leq$ M$_*$/(10$^{10}$ M$_\odot$) $\leq$ 1.4 and 1.0 $\leq$ M$_*$/(10$^{10}$ M$_\odot$) $\leq$ 2.8, respectively.  Below this range there were not enough data for a significant fit; above this mass range, at the radii indicated, the data were not well fitted by equation (\ref{eq:zxsfr}) and are indicated by blanks in Table 4.  All the galaxy bins displayed show a trend of increasing metallicity with mass and generally decreasing metallicity with radial size.

\begin{table}[t]\small
\caption{Native Gas Metallicities (Z$_0\times10^3$) as a Function of Stellar Mass and Galaxy Radius.  (For details see text.)}
\vskip0pt\vskip-6pt\vskip0pt
{\vskip-6pt}
\begin{center}
\begin{tabular}{c|rcc|ccc|cc}
M$_*$	&	\multicolumn{3}{c}{    0.10 $\geq$ z $\geq$ 0.07}&		\multicolumn{3}{c}{0.15$\geq$ z $\geq$ 0.10}&\multicolumn{2}{c}{0.3$\geq$ z$\geq$ 0.15}\\
(10$^{10}$M$_\odot$) &r$_{50}$:  Small  & Med. & Large & Small  & Med.  & Large & Med. & Large \\ \hline
0.36	&	0.90	&		&		&		&		&		&				&	\\
0.50	&	1.06	&	1.06	&		&	1.36	&		&		&				&		\\
0.71	&	1.18	&	1.13	&	1.06	&	1.32	&		&		&				&		\\
1.00	&	1.25	&	1.17	&	1.18	&	1.30	&	1.20	&	1.14	&				&		\\
1.41	&	1.30	&	1.24	&	1.24	&	1.39	&	1.28&	1.22	&				&		\\
2.00	&		&		&		&	1.39	&	1.33	&	1.24	&				&		\\
2.82	&		&		&		&	1.37	&	1.33	&	1.30	&				&	1.30	\\
3.98	&		&		&	1.31	&	1.35	&	1.34	&		&			1.41	&	1.29	\\
5.62	&		&		&	1.27	&	1.34	&		&		&			1.31	&	1.31	\\
7.94	&		&		&		&		&	1.37	&	1.30	&			1.36	&	1.31	\\
11.2	&		&		&		&		&		&		&				&	1.37	\\
\\\noalign{\vskip3pt}
\end{tabular}
\end{center}
\label{tab:Z0summary}
\end{table}

The significance of the parameter $\alpha$ remains to be determined.  This becomes apparent if equation (\ref{eq:crux}) is recast in the equivalent form $(SFR) = \epsilon \dot M_i^{3/2}$, where $\epsilon$ is a constant whose units are $(M_{\odot}/yr)^{-1/2}$and $\alpha = \epsilon^{-2/3}$. Seen in this form, $\epsilon$ is readily interpreted as a factor determining star formation efficiency and inversely proportional to the cube root of the mass infall rate.  In section 5, immediately below, we will show how this efficiency relates to pressure regulated star formation in triggered molecular cloud collapse.

Before leaving this section, we should still address the question of mass balance.  Gas falling into a galaxy should either add to the galaxy's gaseous or stellar mass, or add mass to a central black hole, unless an equal amount of mass is ejected from the galaxy explosively or by means of a variety of potential winds.  Equation (\ref{eq:crux}) thus should not be interpreted as meaning that the duration of infall is identical to the duration of star formation.  As discussed in the next several sections, we deduce that the two processes are sequential, and that their respective durations are generally not identical.

\section{A Mean-Field-Theory Toy Model}

To account the findings reported thus far within the larger framework of astrophysics, we next adopt a toy model based on a mean-field-theoretic approach.  

Consider a distant, highly tenuous, cold clump of gas of mean mass density $\langle\rho\rangle$ directly approaching a galaxy of mass $M$.  Let the relative velocity of the clump at large separations be  $V$.  This clump may have been previously captured into a massive halo surrounding the galaxy.  We may follow the motion of a volume element of the clump displaced by an impact parameter $s$ from the point on the clump that will eventually impact on the galaxy's center.    At large distances from the galaxy, the angular momentum of a unit-mass increment of gas about the galaxy's center is $sV$.  Because gas pressures within a cold tenuous clump may be largely neglected, angular momentum conservation tells us that the velocity $v$ of the mass increment impacting on the galaxy at radius $r$ is given by

\begin{equation}
Vs = vr \ ,
\label{eq:angmom}
\end{equation}

and the infall energy of the increment is

\begin{equation}
\frac{v^2}{2} = \frac{V^2}{2} +\frac {MG}{r}\ . 
\label{eq:energy}\end{equation}

Here G is the gravitational constant and, as usual, the two equations deal solely with the dominant azimuthal velocity components while neglecting minor helical velocity twists. 

Within some radial distance $r_m$ from the galaxy's center, infalling matter will be captured through collision with native gas gravitationally anchored to the galaxy. Beyond $r_m$ the galaxy's gaseous component may be too tenuous and/or $M$ may be too low, and capture will cease. Solving the two equations for $s$ and $V$ we obtain a capture cross section for matter falling into the galaxy within radius $r_m$.  

\begin{equation}
\pi s^2  = \pi \left[r_m^2 +\frac{2r_mMG}{V^2}\right]\ .
\label{eq:capturerad}
\end{equation}

Within a radial distance $r_m$ from a galaxy's center, the mean infall rate depends on the mean intergalactic gas density $\langle\rho\rangle$ as
\begin{equation}
\langle\dot M_i\rangle= \pi \langle \rho \rangle s^2 V = \pi \langle \rho \rangle \left[ r_m^2 + \frac{2r_mMG}{V^2}\right]V\ .
\label{eq:mdoti}
\end{equation}  

Every indication given by the low metallicity of SDSS star-forming galaxies is that star formation at least in these galaxies is initiated by infall of low-metallicity gas.  Triggered star formation is generally thought to commence when the pressure on a molecular cloud is sufficiently high \citep{kru2009}.  The pressure $P$ exerted on the galaxy due to infalling gas will vary as a function of radial distance $r$, and will also depend on infall angle, i.e., whether impact of the infalling gas is at a glancing angle or perpendicular to the gas layer it impacts.  But because our data provide only a single star formation rate per galaxy, the parameter of immediate relevance may be some representative pressure $\langle P \rangle$ suitably averaged over radial distances $r$ and angles of incidence $\theta$.  Taking $v$  now to be the infall velocity at radial distance $r$, and $v\langle f(\theta)g(r)\rangle$ a representative perpendicular-impact-velocity component averaged over all radial distances $r$ and directions $\theta$, we can write the pressure for infall onto an area $A$ within radial distance  $r_m$ from a galaxy's center as

\begin{equation}
\langle P \rangle \sim \frac{\dot M_iv\langle f(\theta)g(r)\rangle}{A} = \langle\rho\rangle\frac{\pi r_m^2}{A}\left[1+\frac{2MG}{r_mV^2}\right]^{3/2}V^2\langle f(\theta)g(r)\rangle\sim \frac{\dot M_i^{3/2}V^{1/2}}{Ar_m(\pi \langle\rho\rangle)^{1/2}}\langle f(\theta)g(r)\rangle\ ,
\label{eq:pearly}
\end{equation}

where we have made use of equations (\ref{eq:angmom}), (\ref{eq:energy}) and (\ref{eq:capturerad}) to eliminate $v$.  

An important feature of equation (\ref{eq:pearly}) is that $\langle P \rangle$ is solely a function of the infall rate $\dot M_i$, galaxy mass $M$, and radius $r_m$, as well as two universal parameters characterizing the infalling matter, a representative velocity for clumps traversing the halo, $V$, and a representative clump mass density, $\langle\rho\rangle$.  Although we have not yet derived estimates for $V$ and $\langle\rho\rangle$ on the basis of available data, we will do so in section 6, below.  

Massive star formation is known to occur principally, if not solely, in giant molecular clouds.  Two recent findings regarding these clouds now enter consideration: 

\citet{kru2012} have shown convincingly that star formation rates are directly proportional $\rho_{c\ell}^{3/2}$,where $\rho_{c\ell}$ is the molecular cloud density, in the Galaxy, local disk galaxies and star bursts, and high redshift disks and starbursts.  The universality of this relationship is  documented by their Figure 3, which establishes a single observationally derived ratio of SFR-per-unit-area $\dot\Sigma_*$ to areal-density-divided-by-free-fall-time, $\Sigma/t_{ff}$.  Within factors of order $\pm 3$, this ratio remains constant at $\dot\Sigma_* \sim 10^{-2}\Sigma/t_{ff}$ over a SFR density range of six orders of magnitude.  Here, $\Sigma$ is the areal density, $t_{ff} = \sqrt{3\pi/32 G\rho_{c\ell}^3}$, and $G$ is the gravitational constant.  Two relevant factors to take away from this is that, both on dynamic and observational grounds, $(SFR)\propto \rho_{c\ell}^{3/2}$, and that the observed star forming efficiency is of order 1\%.

A second consideration is that calculations by \citet{juv2011} show theoretical considerations to imply molecular cloud core temperatures to be remarkably constant.  Equilibrium temperatures, at which cosmic ray heating is balanced by spectral line cooling and cooling by dust to which the gas is coupled, are expected to cluster around $T\sim 10$ K for clouds in the density range $n(H_2) = 3\times 10^2$ to $10^5$ cm$^{-3}$.  These considerations are in general accord with early temperature measurements by \citet{sol1987}.  Gas clouds evidently undergo rapid cooling down to temperatures of order 10K, below which the cooling becomes less efficient.  The temperature constancy across such a wide density range suggests that, over sufficiently long periods to permit heat to be radiated away, molecular clouds thus may be isothermally compressed through application of an external pressure. 

We now examine whether gas falling into a galaxy can produce the required pressure to significantly increase the density of a molecular cloud.  Using a virial coefficient defined as $\alpha_{vir} \equiv 5 \sigma_{tot}^2R/M_gG$, where $R$ is the molecular cloud radius,  $M_g$ is its mass, and  $\sigma_{tot}$ is the one-dimensional thermal plus turbulent velocity dispersion within the cloud.  \citet{kru2009} give the mean pressure in a cloud as $P_{c\ell} = 0.7\times 10^5\alpha_{vir}\Sigma_{c\ell}^{'\ 2}\ k_B$ dyn cm$^{-2}$.  Here $\Sigma'_{c\ell}$ is the areal density of a cloud in units of $\Sigma_{c\ell} \sim 85 M_{\odot}$ pc$^{-2}\sim 2\times 10^{-2}$ g cm$^{_2}$, $k_B$ is the Boltzmann constant, and $\alpha_{vir} \sim 2$  independent of galaxy environment, as  found by \citet{bol2008} and revised by \citet{kru2009}.  A value of $\alpha_{vir} = 1$ implies that the cloud is in self-gravitating virial equilibrium, so that the internal turbulent plus thermal pressures approximately balance gravity.  When $\alpha_{vir} \ll 1$, a cloud undergoes free-fall collapse unless magnetic fields support it. 

To exceed a cloud virial coefficient $\alpha_{vir} \sim 1$ an external pressure needs to typically exceed $P\sim 10^{-11}$ dyn cm$^{-2}$.   We can use equation (\ref{eq:pearly}) to determine the conditions under which pressures of this magnitude may be expected. Setting the factor $\langle f(\theta)g(r)\rangle \sim 1$, we then obtain
\begin{equation}
P = 10^{-11}\left(\frac{\dot M_i}{0.5M_{\odot}{\rm /yr}}\right)^{3/2}\left(\frac{V}{170\ {\rm km/s}}\right)^{1/2}\left(\frac{13\ {\rm kpc}^2}{A}\right)\left(\frac{3.3\ {\rm kpc}}{r_m}\right)\left(\frac{5\times 10^{-28}\ {\rm g/cm}^3}{\rho}\right)^{1/2} {\rm dyne/cm}^2\ ,
\label{eq:pestimate}
\end{equation}
where we have anticipated physical parameters we will justify in section 6 below. The point to recognize, however, is that an infall rate of $\sim 0.5 M_{\odot}$ per year converging on a galactic area  smaller than $\sim 13$ kpc$^2$ will generate a sufficiently high pressure to compress typical galactic clouds and thus trigger star formation.

The triggering process no doubt is complex.  Early calculations of supersonic infall by \citet{har1987} showed that, despite a succession of layers of highly or partially ionized gases, and neutral components separating the tenuous, high-velocity impacting gas from a cold, relatively dense neutral layer in a galaxy's plane, the pressure $P$ remains constant throughout such a layered stack.  Thus the pressure exerted on a galaxy's impact area, if large compared to the scale height of the gaseous component, is also the compressive pressure on a cold cloud in the galaxy's plane.  Recent studies devoted to infall into massive cluster haloes rather than single galaxies are based on similar hydrodynamic considerations \citep{bir2011}.  Roughly similar approaches have been undertaken by \citet{gig2011}.  While highly interesting, these studies still await observational confirmation, which may take some years to obtain.   The connection to the infall of clumps into haloes, and their ultimate impact onto the individual galaxies observed in the SDSS survey, will be discussed below, in sections 7 and 8.

Supersonic infall creates turbulence that intersperses infalling gas with cold neutral components and leads both to compression and eventual diffusive mixing. The rapid cooling of molecular gases, implied by the uniformly low 10 K temperatures pointed out earlier, ultimately also dissipates turbulent structures, leaving the cold gas compressed, at the same temperature $T\sim 10$ K it had before infall, and with a sufficiently low virial coefficient to collapse.  The compression ratio $\Psi$ in this adiabatic sequence is $\gtrsim 4$; its value will depend in part on the ratio of heat capacities $\gamma = c_p/c_v$ in the various turbulent gaseous strata, and in part on the rate of conversion of atomic into molecular hydrogen. Even a value of $\Psi =4$, however, will diminish the free-fall collapse time by a factor of $\sqrt 8$, and thereby greatly accelerate, even if it does not initiate, collapse and star formation.  An acceleration by a factor of 3 in a pre-stellar free-fall collapse, which otherwise might require $\gtrsim 10^7$ yr is significant. Galactic rotational periods are of order $10^8$ yr, so that impact of extragalactic matter on a particular area of a galaxy, will generally produce pressure on this area for no more than $\sim 10^7$ yr.  Over longer periods, the galaxy's rotation gradually moves an impacted area out of the infalling stream that subsequently may fall on neighboring areas.  
 
In summary, clumps of intergalactic matter in haloes, in the density range and with a mass-infall rate indicated by equation (\ref{eq:pestimate}), appear able to both shock-compress and turbulently mix with native cold gases in a galaxy, in times that suffice for the gas to cool to $\sim 10$ K, dissipate the turbulence, and collapse in free-fall over times of order 10$^7$ yr.  

Relevant to the discussion also is work by \citet{elm1993}, which lays the theoretical foundation linking the pressure of the ISM, and the ambient UV radiation field to the resulting fraction of hydrogen in its molecular form, predicting a nearly linear relationship, R$_{mol}\propto$P$^{1.2}$ where R$_{mol}$ represents the fraction of hydrogen in molecular form.  Recent work by other authors have refined this prediction based on empirical observations and simulations, lowering the power law index to $\sim$0.9 \citep{won2002,bli2006,rob2008,ler2008}.

None of this is meant to deny that supernovae trigger star formation as well. The observational evidence for their {\it propagation} of star formation is clear.  But the strong correlation of low metallicity with star formation in SDSS galaxies suggests that star formation is initiated by infall; once initiated, star formation is then propagated further by supernova action.  Hydrodynamic instabilities have a role as well, but the additional pressures exerted by infall and/or explosive ejecta can compress and accelerate the collapse of unstable neighboring regions to give rise to observed O and B star associations.  In addition, although it is by now well established that regions of star formation line spiral arms, the limited spectroscopic aperture of the present study focuses attention on the central portions of galaxies, where spiral arms become less distinct.  Our data thus lack observations that could meaningfully add to a discussion of star formation in spiral arms.

Finally, we note that \citet{ell2008} originally demonstrated a correlation of higher metallicity with lower radii in galaxies of identical mass, and a similar correlation of higher metallicity with lower star formation rates.  Equation (\ref{eq:mdoti}) provides a physical explanation of why infall rates should diminish with diminishing radius; and equation ({\ref{eq:zxmass1}) shows why lower infall rates diminish dilution of the intrinsic metallicities of native gas clouds and thus lead to the observed higher metallicities.  Equation (\ref{eq:crux}) further shows that diminishing infall rates result in diminishing star formation.  

It may be useful, at this point, to still mention that the relation between ${\rm (SFR)} \propto \dot M_i^{3/2}$ in equation  (\ref{eq:crux}) has a superficial resemblance to the Schmidt-Kennicutt (SK) law, $\Sigma_{(SFR)}\propto \Sigma_{gas}^{1.4}$ \citep{ken1998_b}.  Roughly equally good fits to the metallicity dilution could be found with ${\rm (SFR)} \propto \dot M_i^{1.4}$. The SK law, however, deals with the areal density of the native gas in a galaxy, whereas $\dot M_i^{3/2}$ refers to the pressure infalling gas may exert.

We next turn to a justification of the physical parameters given in equation (\ref{eq:pestimate}).

\section{Representative Values of $s$ and $V$}

The wealth of available SDSS data permits us to derive representative values of $s$, $V$ and $\langle\rho\rangle$ by means of relations  (\ref{eq:capturerad}), (\ref{eq:mdoti}), and (\ref{eq:crux}).  We obtain these values for a sampling of galaxies on the galaxy main sequence, i.e., for the most populous and typical galaxy sample available in our data.   We define our observed main sequence as the straight line in log-log space: Log(SFR)=1.12$\times$Log(M$_*$)-11.74.  Along this straight line we have picked five sets of values of galaxy mass and SFR more or less equidistant along the most populated portion of our main sequence.  At each SFR/M$_*$ position we extract a set of sources within a small Log(SFR) / Log(M$_*$) range of $\pm$0.02 and take the median Petrosian half-light radius to be representative of the entire set.  The five sets are summarized in Table \ref{tab:smlattributes}.

We do not know precisely where the infalling material is stopped through interaction with gas in the galaxy, or exactly what fraction of the mass within that radius is stellar, as contrasted to dark matter.  We assume in the equations below that infalling mass is stopped at around two Petrosian radii, and that the enclosed mass within that radius is composed largely of stellar and dark matter with a total mass of $\sim 2\times$M$_*$.  This is roughly consistent with data published by \citet{lei2011b}, who plotted the distribution of stellar mass and total gravitational mass in lensing galaxies (see \citet{lei2011b} Figure 5).  Their total enclosed mass, which they term {\it lensing mass}, plotted as a function of effective, i.e., half-light radius, varies with galaxy stellar mass, but for the range of masses we consider, their ratio of total mass to stellar mass, enclosed within two half-light radii, is roughly 2 to $3M_*$, with sizable error bars.

\begin{table}[t]
\caption{SFRs, Masses, and Radii for Typical Main Sequence Galaxies}
\vskip0pt\vskip-6pt\vskip0pt
{\vskip-6pt}
\begin{center}
\begin{tabular}{ccccccc} \hline\noalign{\vskip3pt}
Galaxy & (SFR)& (SFR)$^{2/3}$ & M$_*$ &$r_{50}$ & \# & $\langle\rho\rangle/\alpha$ at V=170 km/s \\
Set & M$_\odot$/yr & (M$_\odot$/yr)$^{2/3}$ & M$_\odot$ & kpc & sources & (g cm$^{-3})\times$(s g$^{-1}$)$^{1/3}$\\
\noalign{\vskip3pt}\hline\noalign{\vskip3pt}
1 & 0.209 &0.352 & $0.71\times 10^{10}$ & 4.2 & 62 & 1.0$\times 10^{-36}$ \\
2 & 0.417 &0.558 & $1.3\times 10^{10}$ & 4.1 & 141 & 1.3$\times 10^{-36}$ \\
3 & 0.813 &0.871 & $2.4\times 10^{10}$ & 4.4 & 167 & 1.3$\times 10^{-36}$ \\
4 & 1.51 &1.32 & $4.0\times 10^{10}$ & 4.9� & 127 & 1.3$\times 10^{-36}$ \\
5 & 3.16 &2.15 & $7.9\times 10^{10}$ & 5.6� & 64 & 1.0$\times 10^{-36}$ \\

\\\noalign{\vskip3pt}\hline
\end{tabular}
\end{center}
\label{tab:smlattributes}
\end{table}  
We next solve for $\rho$ and $\alpha$ using equations (\ref{eq:capturerad}), (\ref{eq:mdoti}), and (\ref{eq:crux}), finding

\begin{equation}
\langle\rho\rangle/\alpha=\frac{1}{\pi V}\frac{SFR^{2/3}}{r^2} \left(1+\frac{2 M G}{r V^2}\right)^{-1}
\label{eq:rho}\ .
\end{equation}

Based on the assumptions that V has some common characteristic value and $\langle\rho\rangle$ is also more or less constant when averaged over infall rates on main sequence galaxies, we consider a range of V values and examine the resulting $\langle\rho\rangle/\alpha$ values for each of our five sets of sources.  Due to scatter in the data, some difference in the resulting $\langle\rho\rangle/\alpha$ values for our five sets is inevitable.  In figure \ref{fig:minrho}, however, we show that there is a distinct minimum in this scatter at a V value of 170 km/s.  Specifically, this is where the standard deviation of $\langle\rho\rangle/\alpha$ among the five source sets normalized by the mean value, reaches a minimum.  The corresponding mean $\langle\rho\rangle/\alpha$ value is 1.2$\times$10$^{-36}$ (g cm$^{-3}$)(s g$^{-1}$)$^{1/3}$\ .

\begin{figure}[h!]
	\centering
		\includegraphics[width=.8\textwidth,angle=90]{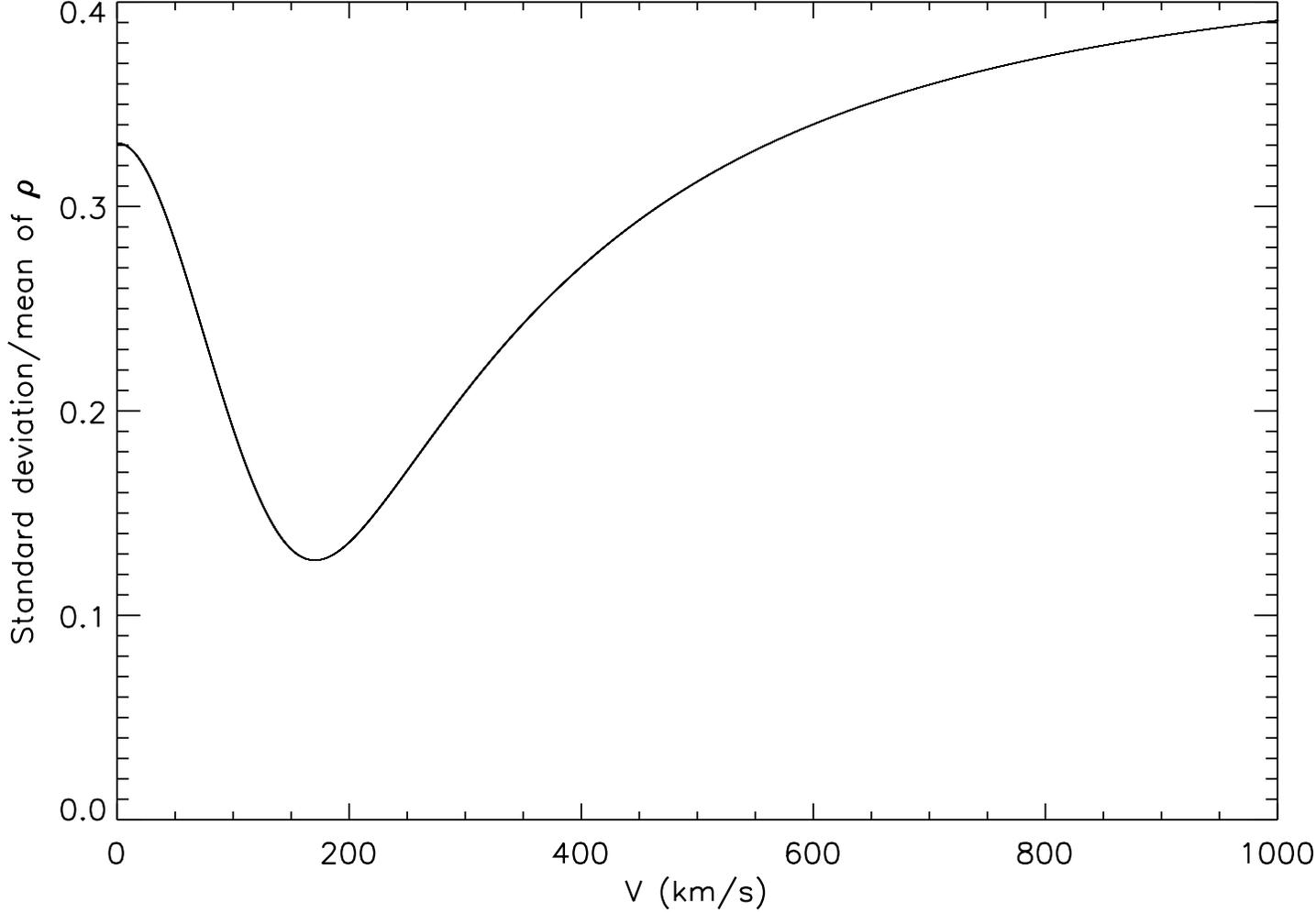}
	\caption{Scatter of $\rho/\alpha$ as a function of V}
	\label{fig:minrho}
\end{figure}

That the infall velocity at large distances, $V$ is as high as $\sim 170$ km s$^{-1}$ may be an indication that gas flowing into a halo containing a group of galaxies acquires this high velocity on falling into the halo, and then impinges on the galaxies without significant further acceleration.  For a halo with radius $\sim 300$ kpc and mass $\sim10^{12} M_{\odot}$, infall velocities of this order would be expected; a more precise value would require knowing the radial distribution of the dark matter within the halo.  Velocities of this order may thus be compared to those obtained from simulations of infall into haloes such as those recently calculated by \citet{kly2011}.

Our estimate of an approach velocity of 170 km s$^{-1}$ is in rough agreement with estimates of $V$ based on the dispersion of line of sight velocities of galaxies in groups and poor clusters, with a range of radial (line-of-sight) velocity dispersion of 100-500 km s$^{-1}$ and a median radial velocity dispersion of $\sim 250$ km s$^{-1}$ \citep{all1999}. 

Adopting a representative mutual approach velocity $V \sim 170$ km s$^{-1}$ for present purposes and setting $\alpha (SFR)^{2/3}$ equal to the mass infall rate $\dot M_i$,we solve for $\langle\rho\rangle$ with the help of equations (\ref{eq:mdoti}) and (\ref{eq:crux}), by writing 
\begin{equation}
\langle\rho\rangle = \frac{\dot M_i}{\pi s^2 V} = \frac{\alpha(SFR)^{2/3}}{\pi s^2V}\ .
\end{equation}

With the parameters given for the five galaxies in Table \ref{tab:smlattributes}, we obtain a mean mass density  $\langle\rho\rangle = (\frac{\alpha}{(M_\odot/yr)^{1/3}})4.8\times 10^{-28}$ g cm$^{-3}$.

In section 7, below, we seek a value of $\alpha$ consistent with available observations in order to obtain an actual estimate of mean clump density $\langle\rho\rangle$.

\section{Difference Between Infall and Star Formation Rates, $\alpha$}

In section 4, we found  that $\alpha \tau_i M_* / M_g$ has a median value of  $\sim 2.3\times 10^{9}$ (yr$^2$ M$_\odot$)$^{1/3}$.  For a typical galaxy with $M_g/M_*= 5\%$, this yields $\alpha\tau_i\sim 1.15\times10^8$ (yr$^2$ M$_\odot$)$^{1/3}$.  We now return to our example of Figure \ref{fig:sampledilution}, where we dealt with galaxies of mass $0.5\times 10^{10}M_{\odot}$.  It will be useful to recast equation (\ref{eq:zxmass1}) in the form

\begin{equation}
\dot M_i\tau_i = \alpha SFR^{2/3} \tau_i = \frac{Z_0-Z_x}{Z_x-Z_i}M_g.
\label{eq:zxsfr2}
\end{equation}

Looking at the metallicities of the most vigorously star forming samples in Figure \ref{fig:sampledilution}, we find that the infalling mass of gas corresponds roughly to 57\% of $M_g$.  For $M_g/M_*=0.05$ we then obtain an infall time of $ \tau_i\sim 1.4\times10^8 M_{\odot}/(\alpha (SFR)^{2/3})\sim 1\times 10^8/\left(\frac{\alpha}{M_\odot^{1/3} yr^{-1/3}}\right)$ yr.  A value of $\alpha = 10$, would reduce the required infall time to $\sim 10^7$ yr, comparable to the time that gas falling onto a limited portion of a galaxy would impinge there before the galaxy's rotation moved neighboring regions in the galaxy into  the infalling stream.

An alternative estimate of $\alpha$ is obtained from parameters derived in section 5 that would produce a pressure sufficiently high to trigger or accelerate free-fall collapse.   For $P\sim (\dot M_i v/A) \sim \alpha(SFR)^{2/3} v/A$ and an infall velocity, as given by equation (\ref{eq:energy}), $v = (s/r)V\sim \sqrt3V\sim 294$ km s$^{-1}$, the infall pressure $P(1M_{\odot}\ {\rm yr}^{-1}$ kpc$^{-1})  \sim 1.3\times 10^{-10}$ dyn cm$^{-2}$, where we have taken $V = 170$ km s$^{-1}$.  Exceeding the internal molecular cloud pressure then would require $\alpha \geq 1$.}
 
Another estimate of $\alpha$ may be suggestive:  Spectra have by now been obtained for approximately $6\times10^5$ SDSS galaxies in our redshift range.  Roughly 17\% of these have indications of star formation and are included in our tables.   If 17\% of the galaxies have measurable star formation rates, 17\% of a typical halo volume should be filled with clumps of density $\langle\rho\rangle = 4.8\times 10^{-28}\left(\frac{\alpha}{M_\odot^{1/3} yr^{-1/3}}\right)$ g cm$^{-3}$, and the gaseous mass density of the haloes should be $8.2\times 10^{-29}\left(\frac{\alpha}{M_\odot^{1/3} yr^{-1/3}}\right)$ g cm$^{-3}$.  For the above-cited halo of radius 300 kpc, the total clump mass would then be $\sim 2.7\times 10^{44}\left(\frac{\alpha}{M_\odot^{1/3} yr^{-1/3}}\right)$ g, or $13.7\times 10^{10}(\frac{\alpha}{M_\odot yr^{-1}}) M_{\odot}$.  This suggests that $\alpha$ should not greatly exceed $\alpha \sim 1\ {\rm to}\ 3 $; otherwise the fraction of  clump mass in the halo would begin to approach that of the total halo mass estimated above as $10^{12} M_{\odot}$.  

In summary, it is possible to defend $\alpha$ values in the range of 1-10 $(M_{\odot}\ {\rm  yr}^{-1})^{1/3}$, representing the difference between values of  $(SFR)^{2/3}$ and $\dot M_i$. This also appears roughly compatible with respective durations of  H$\alpha$ emission from HII regions around massive stars formed in the collapse of a giant molecular cloud and the duration of gas infall onto a particular spot of a rotating galaxy. 

Ultimately, however, the value of $\alpha$ will have to be established through observations.  Extragalactic clumps of gas with mass densities of $\sim 3\times 10^{-27}$ g cm$^{-3}$ and $\alpha$ values in the range of 1 to 10 $(M_{\odot}\ {\rm yr}^{-1})^{1/3}$, can probably not be detected with currently available instrumentation, but should be searched for, once those capabilities become available.

Values of $\alpha \gtrsim 1$ to 10 in the indicated units also make sense when we recall that $\alpha = \epsilon^{-2/3}$,  as discussed in section 4.  As a coefficient for star formation efficiency $\epsilon \dot M_i^{1/2}$, one can only expect $\epsilon$ to exhibit values $\epsilon\leq 1$.   If $\alpha$ is as high as 10, $\epsilon$ would have to be of order $0.03$.  \citet{kru2012} estimate star formation rates of $SFR = f(H_2) \epsilon_{ff} [32G\rho^3/3\pi]^{1/2}$, where $f(H_2)$ is the fraction of gas in the form of molecular hydrogen.  They estimate the free-fall efficiency $f(H_2)\epsilon_{ff}$ to be of order 1\%.  Within reasonable uncertainties, this suggests that an estimate of $\epsilon \sim 0.03$, and thus $\alpha\sim 10$ also lies within an acceptable range.

\section{Summary and Discussion}  

We have shown that for a large fraction of the galaxies whose central regions were observed in the Sloan Digital Sky Survey, the observed gas metallicities, $Z_x$, are consistent with (i) a quantitative physical relation for star formation through episodic infall of gas devoid of heavy elements; (ii) thorough mixing of infalling and native gas before onset of star formation; (iii) a star formation rate proportional to the 3/2 power of the infalling mass rate, $\dot M_i$; and (iv) intermittent quiescent phases devoid of star formation during which the native gas in a galaxy exhibits a characteristic elevated gas metallicity, $Z_0$, dependent on galaxy mass, $M_*$, and a characteristic mean ratio of stellar mass to native mass of gas, $M_g$.  Most if not all our SDSS star-forming galaxies with  $M_* < 2.0\times 10^{10}M_{\odot}$, and many with $M_* \geq 2.0\times 10^{10}M_{\odot}$ and large radii appear fed by infall.   Sets of galaxies not obeying these relations invariably have masses $M_*\geq2.0\times 10^{10}M_{\odot}$ and exhibit high metallicities at high star formation rates, suggesting that for these galaxies star formation independent of mass infall plays a significant role. 

These features are largely reproduced by a mean-field-theory toy model for the physics of infall, which accounts for the $(SFR)\propto \dot M_i^{3/2}$ relation in terms of the pressure infalling gas exerts on native gas in a galaxy to trigger, or at least accelerate, star formation.  The model also permits us to estimate the mean densities and velocities of clumps of baryonic matter traversing the dark matter halos in which the SDSS galaxies may be embedded.  

This is as much conformance to our data as can be expected from a mean-field-theory toy model, which of necessity disregards variations in velocity $V$, variations in clump densities $\rho$, and selection effects that favor detection of larger, more massive galaxies at higher redshifts, and smaller, less massive and less luminous galaxies at lower redshifts.  But it suggests that the basic features of the model capture the essence of population distributions and metallicities found in SDSS star forming galaxies. 

A number of papers published in recent years have discussed the absolute value of chemical  enrichment in galaxies \citep{bro2007,kob2007,fin2008};  modeling the processes involved requires consideration of the chemical constituents of supernova ejecta, their ejection velocities, the gravitational potentials of the parent galaxies, and thus calculations on the rate of retention of metals in galaxies of different types and masses.  As pointed out by \citet{gen2009}, past merger history can also be an important factor.   A number of authors, most recently \citet{lei2011} have also considered the extent to which winds from evolved stars may generate sufficient gas to keep up star formation within quiescent galaxies at a low level on a continual basis, particularly in massive galaxies.  While of great interest, a detailed understanding of these effects will require calculations and detailed modeling beyond the scope of the present paper, as do recent attempts by \citet{bou2010} and \citet{shi2011} to model SFRs through new star-formation laws.  Much of this theoretical work is still tentative and will need to be tested by accumulating observational evidence. Some theoretical predictions could potentially be tested against the quantitative metallicity relations brought out in the present paper.

Of particular interest to such studies may be the upper limits to the metallicity of infalling material that our study indicates, $\sim 0.125\times 10^{-3}\pm 0.125\times 10^{-3}$ N(O)/N(H).  This appears to be a useful constraint on the metallicity of material transiting galaxy halos and falling into the galaxies.

\acknowledgements{
Support for this work was provided by NASA through an award issued by JPL/Caltech.  We wish to thank Dr. F. Mannucci for clarifying for us some of the steps involved in his analyses of gas metallicities in SDSS galaxies. We also appreciate incisive critical comments from Chris McKee and from Bruce Elmegreen, both of which were very helpful. One of us also benefitted from a long and fruitful conversation with Simon Lilly.  We similarly appreciate the many recommendations of an anonymous referee.  
}

\bibliography{mypsrvlb}

\begin{thebibliography}{55}
\expandafter\ifx\csname natexlab\endcsname\relax\def\natexlab#1{#1}\fi

\bibitem[{{Aihara} {et~al.}(2011){Aihara}, {Allende Prieto}, {An}, {Anderson},
  {Aubourg}, {Balbinot}, {Beers}, {Berlind}, {Bickerton}, {Bizyaev}, {Blanton},
  {Bochanski}, {Bolton}, {Bovy}, {Brandt}, {Brinkmann}, {Brown}, {Brownstein},
  {Busca}, {Campbell}, {Carr}, {Chen}, {Chiappini}, {Comparat}, {Connolly},
  {Cortes}, {Croft}, {Cuesta}, {da Costa}, {Davenport}, {Dawson}, {Dhital},
  {Ealet}, {Ebelke}, {Edmondson}, {Eisenstein}, {Escoffier}, {Esposito},
  {Evans}, {Fan}, {Femen{\'{\i}}a Castell{\'a}}, {Font-Ribera}, {Frinchaboy},
  {Ge}, {Gillespie}, {Gilmore}, {Gonz{\'a}lez Hern{\'a}ndez}, {Gott}, {Gould},
  {Grebel}, {Gunn}, {Hamilton}, {Harding}, {Harris}, {Hawley}, {Hearty}, {Ho},
  {Hogg}, {Holtzman}, {Honscheid}, {Inada}, {Ivans}, {Jiang}, {Johnson},
  {Jordan}, {Jordan}, {Kazin}, {Kirkby}, {Klaene}, {Knapp}, {Kneib},
  {Kochanek}, {Koesterke}, {Kollmeier}, {Kron}, {Lampeitl}, {Lang}, {Le Goff},
  {Lee}, {Lin}, {Long}, {Loomis}, {Lucatello}, {Lundgren}, {Lupton}, {Ma},
  {MacDonald}, {Mahadevan}, {Maia}, {Makler}, {Malanushenko}, {Malanushenko},
  {Mandelbaum}, {Maraston}, {Margala}, {Masters}, {McBride}, {McGehee},
  {McGreer}, {M{\'e}nard}, {Miralda-Escud{\'e}}, {Morrison}, {Mullally},
  {Muna}, {Munn}, {Murayama}, {Myers}, {Naugle}, {Fausti Neto}, {Cuong Nguyen},
  {Nichol}, {O'Connell}, {Ogando}, {Olmstead}, {Oravetz}, {Padmanabhan},
  {Palanque-Delabrouille}, {Pan}, {Pandey}, {P{\^a}ris}, {Percival},
  {Petitjean}, {Pfaffenberger}, {Pforr}, {Phleps}, {Pichon}, {Pieri}, {Prada},
  {Price-Whelan}, {Raddick}, {Ramos}, {Reyl{\'e}}, {Rich}, {Richards}, {Rix},
  {Robin}, {Rocha-Pinto}, {Rockosi}, {Roe}, {Rollinde}, {Ross}, {Ross},
  {Rossetto}, {S{\'a}nchez}, {Sayres}, {Schlegel}, {Schlesinger}, {Schmidt},
  {Schneider}, {Sheldon}, {Shu}, {Simmerer}, {Simmons}, {Sivarani}, {Snedden},
  {Sobeck}, {Steinmetz}, {Strauss}, {Szalay}, {Tanaka}, {Thakar}, {Thomas},
  {Tinker}, {Tofflemire}, {Tojeiro}, {Tremonti}, {Vandenberg}, {Vargas
  Maga{\~n}a}, {Verde}, {Vogt}, {Wake}, {Wang}, {Weaver}, {Weinberg}, {White},
  {White}, {Yanny}, {Yasuda}, {Yeche}, \& {Zehavi}}]{aih2011}
{Aihara}, H., {Allende Prieto}, C., {An}, D., {Anderson}, S.~F., {Aubourg},
  {\'E}., {Balbinot}, E., {Beers}, T.~C., {Berlind}, A.~A., {Bickerton}, S.~J.,
  {Bizyaev}, D., {Blanton}, M.~R., {Bochanski}, J.~J., {Bolton}, A.~S., {Bovy},
  J., {Brandt}, W.~N., {Brinkmann}, J., {Brown}, P.~J., {Brownstein}, J.~R.,
  {Busca}, N.~G., {Campbell}, H., {Carr}, M.~A., {Chen}, Y., {Chiappini}, C.,
  {Comparat}, J., {Connolly}, N., {Cortes}, M., {Croft}, R.~A.~C., {Cuesta},
  A.~J., {da Costa}, L.~N., {Davenport}, J.~R.~A., {Dawson}, K., {Dhital}, S.,
  {Ealet}, A., {Ebelke}, G.~L., {Edmondson}, E.~M., {Eisenstein}, D.~J.,
  {Escoffier}, S., {Esposito}, M., {Evans}, M.~L., {Fan}, X., {Femen{\'{\i}}a
  Castell{\'a}}, B., {Font-Ribera}, A., {Frinchaboy}, P.~M., {Ge}, J.,
  {Gillespie}, B.~A., {Gilmore}, G., {Gonz{\'a}lez Hern{\'a}ndez}, J.~I.,
  {Gott}, J.~R., {Gould}, A., {Grebel}, E.~K., {Gunn}, J.~E., {Hamilton},
  J.-C., {Harding}, P., {Harris}, D.~W., {Hawley}, S.~L., {Hearty}, F.~R.,
  {Ho}, S., {Hogg}, D.~W., {Holtzman}, J.~A., {Honscheid}, K., {Inada}, N.,
  {Ivans}, I.~I., {Jiang}, L., {Johnson}, J.~A., {Jordan}, C., {Jordan}, W.~P.,
  {Kazin}, E.~A., {Kirkby}, D., {Klaene}, M.~A., {Knapp}, G.~R., {Kneib},
  J.-P., {Kochanek}, C.~S., {Koesterke}, L., {Kollmeier}, J.~A., {Kron}, R.~G.,
  {Lampeitl}, H., {Lang}, D., {Le Goff}, J.-M., {Lee}, Y.~S., {Lin}, Y.-T.,
  {Long}, D.~C., {Loomis}, C.~P., {Lucatello}, S., {Lundgren}, B., {Lupton},
  R.~H., {Ma}, Z., {MacDonald}, N., {Mahadevan}, S., {Maia}, M.~A.~G.,
  {Makler}, M., {Malanushenko}, E., {Malanushenko}, V., {Mandelbaum}, R.,
  {Maraston}, C., {Margala}, D., {Masters}, K.~L., {McBride}, C.~K., {McGehee},
  P.~M., {McGreer}, I.~D., {M{\'e}nard}, B., {Miralda-Escud{\'e}}, J.,
  {Morrison}, H.~L., {Mullally}, F., {Muna}, D., {Munn}, J.~A., {Murayama}, H.,
  {Myers}, A.~D., {Naugle}, T., {Fausti Neto}, A., {Cuong Nguyen}, D.,
  {Nichol}, R.~C., {O'Connell}, R.~W., {Ogando}, R.~L.~C., {Olmstead}, M.~D.,
  {Oravetz}, D.~J., {Padmanabhan}, N., {Palanque-Delabrouille}, N., {Pan}, K.,
  {Pandey}, P., {P{\^a}ris}, I., {Percival}, W.~J., {Petitjean}, P.,
  {Pfaffenberger}, R., {Pforr}, J., {Phleps}, S., {Pichon}, C., {Pieri}, M.~M.,
  {Prada}, F., {Price-Whelan}, A.~M., {Raddick}, M.~J., {Ramos}, B.~H.~F.,
  {Reyl{\'e}}, C., {Rich}, J., {Richards}, G.~T., {Rix}, H.-W., {Robin}, A.~C.,
  {Rocha-Pinto}, H.~J., {Rockosi}, C.~M., {Roe}, N.~A., {Rollinde}, E., {Ross},
  A.~J., {Ross}, N.~P., {Rossetto}, B.~M., {S{\'a}nchez}, A.~G., {Sayres}, C.,
  {Schlegel}, D.~J., {Schlesinger}, K.~J., {Schmidt}, S.~J., {Schneider},
  D.~P., {Sheldon}, E., {Shu}, Y., {Simmerer}, J., {Simmons}, A.~E.,
  {Sivarani}, T., {Snedden}, S.~A., {Sobeck}, J.~S., {Steinmetz}, M.,
  {Strauss}, M.~A., {Szalay}, A.~S., {Tanaka}, M., {Thakar}, A.~R., {Thomas},
  D., {Tinker}, J.~L., {Tofflemire}, B.~M., {Tojeiro}, R., {Tremonti}, C.~A.,
  {Vandenberg}, J., {Vargas Maga{\~n}a}, M., {Verde}, L., {Vogt}, N.~P.,
  {Wake}, D.~A., {Wang}, J., {Weaver}, B.~A., {Weinberg}, D.~H., {White}, M.,
  {White}, S.~D.~M., {Yanny}, B., {Yasuda}, N., {Yeche}, C., \& {Zehavi}, I.
  2011, \apjs, 193, 29

\bibitem[{Allen \& Cox(1999)}]{all1999}
Allen, C., \& Cox, A. 1999, Allen's Astrophysical Quantities, 4th ed.
  (Springer)

\bibitem[{{Birnboim} \& {Dekel}(2011)}]{bir2011}
{Birnboim}, Y., \& {Dekel}, A. 2011, \mnras, 415, 2566

\bibitem[{{Blitz} \& {Rosolowsky}(2006)}]{bli2006}
{Blitz}, L., \& {Rosolowsky}, E. 2006, \apj, 650, 933

\bibitem[{{Bolatto} {et~al.}(2008){Bolatto}, {Leroy}, {Rosolowsky}, {Walter},
  \& {Blitz}}]{bol2008}
{Bolatto}, A.~D., {Leroy}, A.~K., {Rosolowsky}, E., {Walter}, F., \& {Blitz},
  L. 2008, \apj, 686, 948

\bibitem[{{Bouch{\'e}} {et~al.}(2010){Bouch{\'e}}, {Dekel}, {Genzel}, {Genel},
  {Cresci}, {F{\"o}rster Schreiber}, {Shapiro}, {Davies}, \&
  {Tacconi}}]{bou2010}
{Bouch{\'e}}, N., {Dekel}, A., {Genzel}, R., {Genel}, S., {Cresci}, G.,
  {F{\"o}rster Schreiber}, N.~M., {Shapiro}, K.~L., {Davies}, R.~I., \&
  {Tacconi}, L. 2010, \apj, 718, 1001

\bibitem[{{Brooks} {et~al.}(2007){Brooks}, {Governato}, {Booth}, {Willman},
  {Gardner}, {Wadsley}, {Stinson}, \& {Quinn}}]{bro2007}
{Brooks}, A.~M., {Governato}, F., {Booth}, C.~M., {Willman}, B., {Gardner},
  J.~P., {Wadsley}, J., {Stinson}, G., \& {Quinn}, T. 2007, \apjl, 655, L17

\bibitem[{{Calzetti} \& {Kennicutt}(2009)}]{cal2009}
{Calzetti}, D., \& {Kennicutt}, R.~C. 2009, \pasp, 121, 937

\bibitem[{{Cardelli} {et~al.}(1989){Cardelli}, {Clayton}, \&
  {Mathis}}]{car1989}
{Cardelli}, J.~A., {Clayton}, G.~C., \& {Mathis}, J.~S. 1989, \apj, 345, 245

\bibitem[{{Chabrier}(2003)}]{cha2003}
{Chabrier}, G. 2003, \pasp, 115, 763

\bibitem[{Chabrier(2005)}]{cha2005}
Chabrier, G. 2005, The Initial Mass Function 50 Years Later, ed Corelli E.,
  Palla F., \& Zinnecker H

\bibitem[{{Cowie} {et~al.}(2011){Cowie}, {Barger}, \& {Hu}}]{cow2011}
{Cowie}, L.~L., {Barger}, A.~J., \& {Hu}, E.~M. 2011, \apj, 738, 136

\bibitem[{{Dalcanton}(2007)}]{dal2007}
{Dalcanton}, J.~J. 2007, \apj, 658, 941

\bibitem[{{Dav{\'e}} {et~al.}(2011{\natexlab{a}}){Dav{\'e}}, {Finlator}, \&
  {Oppenheimer}}]{dav2011}
{Dav{\'e}}, R., {Finlator}, K., \& {Oppenheimer}, B.~D. 2011{\natexlab{a}},
  \mnras, 416, 1354

\bibitem[{{Dav{\'e}} {et~al.}(2011{\natexlab{b}}){Dav{\'e}}, {Oppenheimer}, \&
  {Finlator}}]{dav2011b}
{Dav{\'e}}, R., {Oppenheimer}, B.~D., \& {Finlator}, K. 2011{\natexlab{b}},
  \mnras, 415, 11

\bibitem[{{Ellison} {et~al.}(2008){Ellison}, {Patton}, {Simard}, \&
  {McConnachie}}]{ell2008}
{Ellison}, S.~L., {Patton}, D.~R., {Simard}, L., \& {McConnachie}, A.~W. 2008,
  \apjl, 672, L107

\bibitem[{{Elmegreen}(1993)}]{elm1993}
{Elmegreen}, B.~G. 1993, \apj, 411, 170

\bibitem[{{Faucher-Gigu{\`e}re} {et~al.}(2011){Faucher-Gigu{\`e}re}, {Kere{\v
  s}}, \& {Ma}}]{gig2011}
{Faucher-Gigu{\`e}re}, C.-A., {Kere{\v s}}, D., \& {Ma}, C.-P. 2011, \mnras,
  417, 2982

\bibitem[{{Finlator} \& {Dav{\'e}}(2008)}]{fin2008}
{Finlator}, K., \& {Dav{\'e}}, R. 2008, \mnras, 385, 2181

\bibitem[{{Gallazzi} {et~al.}(2005){Gallazzi}, {Charlot}, {Brinchmann},
  {White}, \& {Tremonti}}]{gal2005}
{Gallazzi}, A., {Charlot}, S., {Brinchmann}, J., {White}, S.~D.~M., \&
  {Tremonti}, C.~A. 2005, \mnras, 362, 41

\bibitem[{{Genel} {et~al.}(2009){Genel}, {Genzel}, {Bouch{\'e}}, {Naab}, \&
  {Sternberg}}]{gen2009}
{Genel}, S., {Genzel}, R., {Bouch{\'e}}, N., {Naab}, T., \& {Sternberg}, A.
  2009, \apj, 701, 2002

\bibitem[{{Harwit} {et~al.}(1987){Harwit}, {Houck}, {Soifer}, \&
  {Palumbo}}]{har1987}
{Harwit}, M., {Houck}, J.~R., {Soifer}, B.~T., \& {Palumbo}, G.~G.~C. 1987,
  \apj, 315, 28

\bibitem[{{Juvela} \& {Ysard}(2011)}]{juv2011}
{Juvela}, M., \& {Ysard}, N. 2011, \apj, 739, 63

\bibitem[{{Kauffmann} {et~al.}(2003{\natexlab{a}}){Kauffmann}, {Heckman},
  {Tremonti}, {Brinchmann}, {Charlot}, {White}, {Ridgway}, {Brinkmann},
  {Fukugita}, {Hall}, {Ivezi{\'c}}, {Richards}, \& {Schneider}}]{kau2003}
{Kauffmann}, G., {Heckman}, T.~M., {Tremonti}, C., {Brinchmann}, J., {Charlot},
  S., {White}, S.~D.~M., {Ridgway}, S.~E., {Brinkmann}, J., {Fukugita}, M.,
  {Hall}, P.~B., {Ivezi{\'c}}, {\v Z}., {Richards}, G.~T., \& {Schneider},
  D.~P. 2003{\natexlab{a}}, \mnras, 346, 1055

\bibitem[{{Kauffmann} {et~al.}(2003{\natexlab{b}}){Kauffmann}, {Heckman},
  {White}, {Charlot}, {Tremonti}, {Brinchmann}, {Bruzual}, {Peng}, {Seibert},
  {Bernardi}, {Blanton}, {Brinkmann}, {Castander}, {Cs{\'a}bai}, {Fukugita},
  {Ivezic}, {Munn}, {Nichol}, {Padmanabhan}, {Thakar}, {Weinberg}, \&
  {York}}]{kau2003_b}
{Kauffmann}, G., {Heckman}, T.~M., {White}, S.~D.~M., {Charlot}, S.,
  {Tremonti}, C., {Brinchmann}, J., {Bruzual}, G., {Peng}, E.~W., {Seibert},
  M., {Bernardi}, M., {Blanton}, M., {Brinkmann}, J., {Castander}, F.,
  {Cs{\'a}bai}, I., {Fukugita}, M., {Ivezic}, Z., {Munn}, J.~A., {Nichol},
  R.~C., {Padmanabhan}, N., {Thakar}, A.~R., {Weinberg}, D.~H., \& {York}, D.
  2003{\natexlab{b}}, \mnras, 341, 33

\bibitem[{{Kennicutt}(1998{\natexlab{a}})}]{ken1998}
{Kennicutt}, Jr., R.~C. 1998{\natexlab{a}}, \araa, 36, 189

\bibitem[{{Kennicutt}(1998{\natexlab{b}})}]{ken1998_b}
---. 1998{\natexlab{b}}, \apj, 498, 541

\bibitem[{{Kewley} \& {Ellison}(2008)}]{kew2008}
{Kewley}, L.~J., \& {Ellison}, S.~L. 2008, \apj, 681, 1183

\bibitem[{{Klypin} {et~al.}(2011){Klypin}, {Trujillo-Gomez}, \&
  {Primack}}]{kly2011}
{Klypin}, A.~A., {Trujillo-Gomez}, S., \& {Primack}, J. 2011, \apj, 740, 102

\bibitem[{{Kobayashi} {et~al.}(2007){Kobayashi}, {Springel}, \&
  {White}}]{kob2007}
{Kobayashi}, C., {Springel}, V., \& {White}, S.~D.~M. 2007, \mnras, 376, 1465

\bibitem[{{Kroupa}(2001)}]{kro2001}
{Kroupa}, P. 2001, \mnras, 322, 231

\bibitem[{{Krumholz} {et~al.}(2012){Krumholz}, {Dekel}, \& {McKee}}]{kru2012}
{Krumholz}, M.~R., {Dekel}, A., \& {McKee}, C.~F. 2012, \apj, 745, 69

\bibitem[{{Krumholz} {et~al.}(2009){Krumholz}, {McKee}, \&
  {Tumlinson}}]{kru2009}
{Krumholz}, M.~R., {McKee}, C.~F., \& {Tumlinson}, J. 2009, \apj, 699, 850

\bibitem[{{Lara-L{\'o}pez} {et~al.}(2010){Lara-L{\'o}pez}, {Cepa},
  {Bongiovanni}, {P{\'e}rez Garc{\'{\i}}a}, {Ederoclite}, {Casta{\~n}eda},
  {Fern{\'a}ndez Lorenzo}, {Povi{\'c}}, \& {S{\'a}nchez-Portal}}]{lar2010}
{Lara-L{\'o}pez}, M.~A., {Cepa}, J., {Bongiovanni}, A., {P{\'e}rez
  Garc{\'{\i}}a}, A.~M., {Ederoclite}, A., {Casta{\~n}eda}, H., {Fern{\'a}ndez
  Lorenzo}, M., {Povi{\'c}}, M., \& {S{\'a}nchez-Portal}, M. 2010, \aap, 521,
  L53

\bibitem[{{Lehner} \& {Howk}(2011)}]{leh2011}
{Lehner}, N., \& {Howk}, J.~C. 2011, Science, 334, 955

\bibitem[{{Leier} {et~al.}(2011){Leier}, {Ferreras}, {Saha}, \&
  {Falco}}]{lei2011b}
{Leier}, D., {Ferreras}, I., {Saha}, P., \& {Falco}, E.~E. 2011, \apj, 740, 97

\bibitem[{{Leitner} \& {Kravtsov}(2011)}]{lei2011}
{Leitner}, S.~N., \& {Kravtsov}, A.~V. 2011, \apj, 734, 48

\bibitem[{{Leroy} {et~al.}(2008){Leroy}, {Walter}, {Brinks}, {Bigiel}, {de
  Blok}, {Madore}, \& {Thornley}}]{ler2008}
{Leroy}, A.~K., {Walter}, F., {Brinks}, E., {Bigiel}, F., {de Blok}, W.~J.~G.,
  {Madore}, B., \& {Thornley}, M.~D. 2008, \aj, 136, 2782

\bibitem[{{Madore}(2010)}]{mad2010}
{Madore}, B.~F. 2010, \apjl, 716, L131

\bibitem[{{Maiolino} {et~al.}(2008){Maiolino}, {Nagao}, {Grazian}, {Cocchia},
  {Marconi}, {Mannucci}, {Cimatti}, {Pipino}, {Ballero}, {Calura}, {Chiappini},
  {Fontana}, {Granato}, {Matteucci}, {Pastorini}, {Pentericci}, {Risaliti},
  {Salvati}, \& {Silva}}]{mai2008}
{Maiolino}, R., {Nagao}, T., {Grazian}, A., {Cocchia}, F., {Marconi}, A.,
  {Mannucci}, F., {Cimatti}, A., {Pipino}, A., {Ballero}, S., {Calura}, F.,
  {Chiappini}, C., {Fontana}, A., {Granato}, G.~L., {Matteucci}, F.,
  {Pastorini}, G., {Pentericci}, L., {Risaliti}, G., {Salvati}, M., \& {Silva},
  L. 2008, \aap, 488, 463

\bibitem[{{Mannucci} {et~al.}(2010){Mannucci}, {Cresci}, {Maiolino}, {Marconi},
  \& {Gnerucci}}]{man2010}
{Mannucci}, F., {Cresci}, G., {Maiolino}, R., {Marconi}, A., \& {Gnerucci}, A.
  2010, \mnras, 408, 2115

\bibitem[{{Nagao} {et~al.}(2006){Nagao}, {Maiolino}, \& {Marconi}}]{nag2006}
{Nagao}, T., {Maiolino}, R., \& {Marconi}, A. 2006, \aap, 459, 85

\bibitem[{{Parravano} {et~al.}(2011){Parravano}, {McKee}, \&
  {Hollenbach}}]{par2011}
{Parravano}, A., {McKee}, C.~F., \& {Hollenbach}, D.~J. 2011, \apj, 726, 27

\bibitem[{{Peeples} \& {Shankar}(2011)}]{pee2011}
{Peeples}, M.~S., \& {Shankar}, F. 2011, \mnras, 417, 2962

\bibitem[{{Peng} {et~al.}(2011){Peng}, {Lilly}, {Renzini}, \&
  {Carollo}}]{pen2011}
{Peng}, Y., {Lilly}, S.~J., {Renzini}, A., \& {Carollo}, M. 2011, ArXiv
  e-prints

\bibitem[{{Peng} {et~al.}(2010){Peng}, {Lilly}, {Kova{\v c}}, {Bolzonella},
  {Pozzetti}, {Renzini}, {Zamorani}, {Ilbert}, {Knobel}, {Iovino}, {Maier},
  {Cucciati}, {Tasca}, {Carollo}, {Silverman}, {Kampczyk}, {de Ravel},
  {Sanders}, {Scoville}, {Contini}, {Mainieri}, {Scodeggio}, {Kneib}, {Le
  F{\`e}vre}, {Bardelli}, {Bongiorno}, {Caputi}, {Coppa}, {de la Torre},
  {Franzetti}, {Garilli}, {Lamareille}, {Le Borgne}, {Le Brun}, {Mignoli},
  {Perez Montero}, {Pello}, {Ricciardelli}, {Tanaka}, {Tresse}, {Vergani},
  {Welikala}, {Zucca}, {Oesch}, {Abbas}, {Barnes}, {Bordoloi}, {Bottini},
  {Cappi}, {Cassata}, {Cimatti}, {Fumana}, {Hasinger}, {Koekemoer},
  {Leauthaud}, {Maccagni}, {Marinoni}, {McCracken}, {Memeo}, {Meneux}, {Nair},
  {Porciani}, {Presotto}, \& {Scaramella}}]{pen2010}
{Peng}, Y.-j., {Lilly}, S.~J., {Kova{\v c}}, K., {Bolzonella}, M., {Pozzetti},
  L., {Renzini}, A., {Zamorani}, G., {Ilbert}, O., {Knobel}, C., {Iovino}, A.,
  {Maier}, C., {Cucciati}, O., {Tasca}, L., {Carollo}, C.~M., {Silverman}, J.,
  {Kampczyk}, P., {de Ravel}, L., {Sanders}, D., {Scoville}, N., {Contini}, T.,
  {Mainieri}, V., {Scodeggio}, M., {Kneib}, J.-P., {Le F{\`e}vre}, O.,
  {Bardelli}, S., {Bongiorno}, A., {Caputi}, K., {Coppa}, G., {de la Torre},
  S., {Franzetti}, P., {Garilli}, B., {Lamareille}, F., {Le Borgne}, J.-F., {Le
  Brun}, V., {Mignoli}, M., {Perez Montero}, E., {Pello}, R., {Ricciardelli},
  E., {Tanaka}, M., {Tresse}, L., {Vergani}, D., {Welikala}, N., {Zucca}, E.,
  {Oesch}, P., {Abbas}, U., {Barnes}, L., {Bordoloi}, R., {Bottini}, D.,
  {Cappi}, A., {Cassata}, P., {Cimatti}, A., {Fumana}, M., {Hasinger}, G.,
  {Koekemoer}, A., {Leauthaud}, A., {Maccagni}, D., {Marinoni}, C.,
  {McCracken}, H., {Memeo}, P., {Meneux}, B., {Nair}, P., {Porciani}, C.,
  {Presotto}, V., \& {Scaramella}, R. 2010, \apj, 721, 193

\bibitem[{{Robertson} \& {Kravtsov}(2008)}]{rob2008}
{Robertson}, B.~E., \& {Kravtsov}, A.~V. 2008, \apj, 680, 1083

\bibitem[{{Savaglio} {et~al.}(2005){Savaglio}, {Glazebrook}, {Le Borgne},
  {Juneau}, {Abraham}, {Chen}, {Crampton}, {McCarthy}, {Carlberg}, {Marzke},
  {Roth}, {J{\o}rgensen}, \& {Murowinski}}]{sav2005}
{Savaglio}, S., {Glazebrook}, K., {Le Borgne}, D., {Juneau}, S., {Abraham},
  R.~G., {Chen}, H.-W., {Crampton}, D., {McCarthy}, P.~J., {Carlberg}, R.~G.,
  {Marzke}, R.~O., {Roth}, K., {J{\o}rgensen}, I., \& {Murowinski}, R. 2005,
  \apj, 635, 260

\bibitem[{{Shi} {et~al.}(2011){Shi}, {Helou}, {Yan}, {Armus}, {Wu}, {Papovich},
  \& {Stierwalt}}]{shi2011}
{Shi}, Y., {Helou}, G., {Yan}, L., {Armus}, L., {Wu}, Y., {Papovich}, C., \&
  {Stierwalt}, S. 2011, \apj, 733, 87

\bibitem[{{Solomon} {et~al.}(1987){Solomon}, {Rivolo}, {Barrett}, \&
  {Yahil}}]{sol1987}
{Solomon}, P.~M., {Rivolo}, A.~R., {Barrett}, J., \& {Yahil}, A. 1987, \apj,
  319, 730

\bibitem[{{Tremonti} {et~al.}(2004){Tremonti}, {Heckman}, {Kauffmann},
  {Brinchmann}, {Charlot}, {White}, {Seibert}, {Peng}, {Schlegel}, {Uomoto},
  {Fukugita}, \& {Brinkmann}}]{tre2004}
{Tremonti}, C.~A., {Heckman}, T.~M., {Kauffmann}, G., {Brinchmann}, J.,
  {Charlot}, S., {White}, S.~D.~M., {Seibert}, M., {Peng}, E.~W., {Schlegel},
  D.~J., {Uomoto}, A., {Fukugita}, M., \& {Brinkmann}, J. 2004, \apj, 613, 898

\bibitem[{{Tripp} {et~al.}(2011){Tripp}, {Meiring}, {Prochaska}, {Willmer},
  {Howk}, {Werk}, {Jenkins}, {Bowen}, {Lehner}, {Sembach}, {Thom}, \&
  {Tumlinson}}]{tri2011}
{Tripp}, T.~M., {Meiring}, J.~D., {Prochaska}, J.~X., {Willmer}, C.~N.~A.,
  {Howk}, J.~C., {Werk}, J.~K., {Jenkins}, E.~B., {Bowen}, D.~V., {Lehner}, N.,
  {Sembach}, K.~R., {Thom}, C., \& {Tumlinson}, J. 2011, Science, 334, 952

\bibitem[{{Wong} \& {Blitz}(2002)}]{won2002}
{Wong}, T., \& {Blitz}, L. 2002, \apj, 569, 157

\bibitem[{{Wright}(2006)}]{wri2006}
{Wright}, E.~L. 2006, \pasp, 118, 1711

\bibitem[{{Yates} {et~al.}(2011){Yates}, {Kauffmann}, \& {Guo}}]{yat2011}
{Yates}, R.~M., {Kauffmann}, G., \& {Guo}, Q. 2011, ArXiv e-prints

\end{thebibliography}

\end{document}